\documentclass[pre,eqsecnum,twocolumn, showpacs]{revtex4}
\global\arraycolsep=2pt

\usepackage{latexsym}
\usepackage{amsmath}
\usepackage{graphicx}
\newcommand{\be}{\begin{equation}}
\newcommand{\ee}{\end{equation}}

\begin{document}

\title{Electrodynamic Casimir Effect in a Medium-filled Wedge}

\date{\today}

\author{Iver Brevik}\email{iver.h.brevik@ntnu.no} 
\author{Simen {\AA}. Ellingsen}\email{simen.a.ellingsen@ntnu.no}
\affiliation{Department of Energy and Process Engineering, Norwegian University
of Science and Technology, N-7491 Trondheim, Norway}

\author{Kimball A. Milton}\email{milton@nhn.ou.edu}

\affiliation{Oklahoma Center for High Energy Physics and 
Homer L. Dodge Department of Physics
and Astronomy, The University of Oklahoma, Norman, Oklahoma 73019,
USA}

\begin{abstract}
We re-examine the electrodynamic Casimir effect in a wedge defined
by two perfect conductors making dihedral angle $\alpha=\pi/p$.
This system is analogous to the system defined by a cosmic string.
We consider the wedge region as filled with an azimuthally symmetric
material, with permittivity/permeability $\varepsilon_1,\mu_1$ for
distance from the axis $r<a$, and $\varepsilon_2,\mu_2$ for $r>a$.
The results are closely related to those for a circular-cylindrical
geometry, but with non-integer azimuthal quantum number $mp$.  
Apart from a zero-mode divergence, which may be removed by
choosing periodic boundary conditions on the wedge, and may
be made finite if dispersion is included, we obtain
finite results for the free energy corresponding to changes in $a$
 for the case when the speed of light is the same
inside and outside the radius $a$, and for weak coupling, $|\varepsilon_1-
\varepsilon_2|\ll1$, for purely dielectric media. 
We also consider the radiation produced by the sudden appearance of an 
infinite cosmic string, situated along the cusp line of the pre-existing 
wedge.
\end{abstract}

\maketitle

%%%%%%%%%%%%%%%%%%%%%%%%%%%%%%%%%%%%%%%%%%%%%%%%
%%%%%%%%%%%%%%%% S E C T I O N %%%%%%%%%%%%%%%%%
%%%%%%%%%%%%%%%%%%%%%%%%%%%%%%%%%%%%%%%%%%%%%%%%
\section{Introduction}

Quantum field theory in the wedge geometry continues to attract
interest, especially in connection with the Casimir effect.
Usually it is assumed that the interior region of the wedge is a
vacuum, and that the two plane surfaces $\theta=0$ and
$\theta=\alpha$ ($\alpha$ denotes the opening angle) are perfectly
conducting. The coordinate system is conventionally oriented such
that the $z$ axis coincides with the the singularity axis, i.e.,
the intersection line for the planes. For an introduction to the
wedge model one may consult the book of Mostepanenko and Trunov
\cite{mostepanenko97}. 

The Casimir energy and stress in a wedge geometry was approached already 
in the 1970s \cite{dowker78, deutsch79}. Since that time, various 
embodiments of the wedge with perfectly conducting walls have been treated by 
Brevik and co-workers \cite{brevik96,brevik98, brevik01} and others 
\cite{nesterenko02}. More recently a wedge intercut by a cylindrical shell was 
considered by Nesterenko and collaborators, first for a
semicircular wedge \cite{nest01}, then for arbitrary dihedral angle
\cite{nest03}.
Local Casimir stresses were examined by Saharian and co-workers 
\cite{rezaeian02,saharian07, saharian08}.  Rosa and collaborators 
studied the interaction of an atom with a wedge \cite{mendes08, rosa08}, 
the situation under which the closely related Casimir-Polder force was 
investigated by Sukenik et al.\ some years ago \cite{sukenik93}.  That 
interaction was first worked out by Barton \cite{barton}.

One reason for the interest in the wedge geometry is the
similarity with the formalism encountered in Casimir theory of
systems having circular symmetry. This applies to the case of a
perfectly conducting circular boundary 
\cite{deraad81, godzinsky98, milton99, lambiase99}, as well as
to the case of a dielectric circular boundary 
\cite{cavero05, cavero06, romeo05, romeo06, brevik07}.
Another reason for studying the wedge is the analogy---at least in
a formal sense---with the theory of a cosmic string (cf., for
instance, Ref.~\cite{vilenkin94}, or Ref.~\cite{brevik96}). Let us
briefly elaborate on the last-mentioned point. The line element
outside a cosmic string is, in standard notation,
\begin{equation}
ds^2=-dt^2+dr^2+(1-4GM)^2 r^2 d\theta^2+dz^2, \label{1}
\end{equation}
where $G$ is the gravitational constant and $M$ the string mass
per unit length. This is the geometry of locally flat space, with
a deficit angle $\Phi=8\pi GM$ being removed. Let us introduce the
symbols $\beta$ and $p$ by
\begin{subequations}
\begin{eqnarray}
\beta&=&(1-4GM)^{-1}=(1-\Phi/2\pi)^{-1}, \label{2}
\\
p&=&\pi/\alpha. \label{3}
\end{eqnarray}
\end{subequations}
 Now comparing the
electromagnetic energy-momentum tensor outside the string
\cite{frolov87}
\begin{equation}
\langle
T_{\mu\nu}\rangle=\frac{1}{720\pi^2r^4}(\beta^2+11)(\beta^2-1){\rm
diag}(1,-3,1,1) \label{4}
\end{equation}
with the electromagnetic energy-momentum in the wedge
\cite{dowker78, deutsch79,brevik96}
\begin{equation}
 \langle
T_{\mu\nu}\rangle=\frac{1}{720\pi^2r^4}\left(\frac{\pi^2}{\alpha^2}+11\right)
\left(\frac{\pi^2}{\alpha^2}-1\right) {\rm diag} (1,-3,1,1),
\label{5}
\end{equation}
we see that $\beta$ corresponds to  $p$. Hence the deficit angle
$\Phi$ corresponds to $2\pi-2\alpha$. We shall return to this
analogy later.  Note that the stress tensor diverges at $r=0$, which makes
the definition of a total Casimir energy in these configurations 
problematic. (Possible solutions to this problem were offered by
Khusnutdinov and Bordag \cite{khus99}.)

\begin{figure}[htb]
  \includegraphics[width=2in]{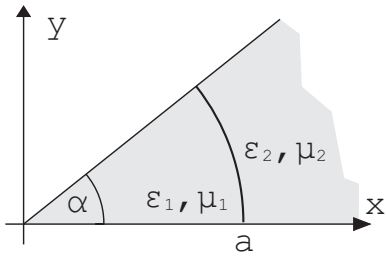}
\caption{The geometry considered in sections \ref{sec_int} and \ref{sec_ext}. 
There is a cylindrical perfectly conducting shell at radius $a$. 
In these sections the indices of refraction are equal,
$n^2=\epsilon_1\mu_1=\epsilon_2\mu_2$.}
  \label{fig_cylinder}
\end{figure}

A particular variant of the wedge model occurs if we introduce a
cylindrical boundary of radius $a$ in the cavity. The situation is
sketched in Fig.~\ref{fig_cylinder}. 
This model has been studied in particular by Nesterenko et al.\ and
by Saharian et al.; 
cf.~Refs.~\cite{nest01,nest03,rezaeian02,saharian07,saharian08} 
with a wealth
of further references therein.  (For example, the fermionic
situation for the circular case was discussed by Bezzera de Mello
et al.~\cite{bdm08}.)
 The model can be looked upon as
being intermediate between that of a conventional wedge, and an
optical fiber. And that brings us to the main theme of the present
paper, namely to study the situation of Fig.~\ref{fig_cylinder}  
in the presence of
a dielectric medium, both in the interior $r<a$ as well as
in the exterior, $r>a$. We designate the two regions by indices 1
and 2. Thus in the interior the refractive index is
$n_1=\sqrt{\epsilon_1 \mu_1}$ with $\epsilon_1$ and $\mu_1$ being
the permittivity and the permeability, whereas in the exterior we
have analogously $n_2=\sqrt{\epsilon_2\mu_2}$. We take all
material quantities $\epsilon_1, \mu_1$ and $\epsilon_2, \mu_2$ to
be constant and nondispersive. The special 
case when the circular boundary is perfectly
conducting is included in the general situation when there is simply a 
dielectric/diamagnetic boundary. The plane
surfaces $\theta=0$ and $\theta=\alpha$ are taken to be perfectly
conducting, as usual.

We begin in the next section by considering the Fourier
decomposition of the TE and TM modes when the circular boundary is
perfectly conducting. This is the simplest case. Then we move on
to give an expression for
 the Casimir energy. The case of a dielectric/diamagnetic
boundary is considered thereafter. The results for the wedge are
in general divergent, not because of the 
divergence associated with the apex of the wedge, which
does not contribute to the outward stress on the circular arc,
but because of the corners where the arc meets the sides of the wedge.
This divergence may be isolated in the azimuthal zero modes, independent
of the angular coordinates.  We propose isolation and removal of this
divergence; alternatively, if the perfectly conducting boundaries at
$\theta=0,\alpha$ are replaced by periodic boundary conditions, these
divergences disappear.  When either of these devices are employed,
we obtain numerical
results for the resulting finite Casimir energy, referring to the boundary
between the two regions, $r<a$ and $r>a$, both for weak and strong 
coupling.  Finally, we exploit the analogy
with a cosmic string to calculate, via the Bogoliubov transformation,
the production of electromagnetic energy associated with a
``sudden" creation of the full wedge situation, as compared with
the initial case of a single-medium filled wedge.

%%%%%%%%%%%%%%%%%%%%%%%%%%%%%%%%%%%%%%%%%%%%%%%%%%%%
%%%%%%%%%%%%%%%%% S E C T I O N %%%%%%%%%%%%%%%%%%%%
%%%%%%%%%%%%%%%%%%%%%%%%%%%%%%%%%%%%%%%%%%%%%%%%%%%%

\section{Zero-point energy in the interior region---Perfectly
conducting arc}\label{sec_int}

As mentioned, we consider an isotropic and homogeneous medium with
 permittivity $\epsilon_1$ and permeability $\mu_1$
enclosed within a wedge region limited by the conducting plane
surfaces $\theta=0$ and $\theta =\alpha$ ($ \leq 2\pi$). In an
$xy$ plane, the  cusp is situated at the origin. We  use
cylindrical coordinates $(r,\theta, z)$. 
%For simplicity we assume
%that the parameter $p$ introduced in Eq.~(\ref{3}) is an integer 
%({\bf CHECK: IS THIS NECESSARY?}).
We employ Heaviside-Lorentz units, and put $\hbar$ and $c$ equal
to unity.

Assume, to begin with, that the wedge is closed by a perfectly
conducting singular arc at $r=a$. We write down the fundamental
modes for stationary electromagnetic modes in the interior wedge,
by invoking the expansions given in Ref.~\cite{stratton41}:
\begin{widetext}
\begin{subequations}
\begin{align}
E_r=&\sum_{m=1}^\infty \left[
-\frac{2k}{\lambda_1}J'_{mp}(\lambda_1 r)a_{m}^i-\frac{2i\mu_1
\omega mp}{\lambda_1^2r}J_{mp}(\lambda_1 r)b_{m}^i\right]F_0\sin
mp\theta, \label{6} \\
E_\theta=&-{\sum_{m=0}^\infty}\left[
\frac{2kmp}{\lambda_1^2r}J_{mp}(\lambda_1 r)a_{m}^i+\frac{2i\mu_1
\omega}{\lambda_1}J'_{mp}(\lambda_1 r)b_{m}^i\right]F_0\cos
mp\theta, \label{7}\\
E_z=&2i\sum_{m=1}^\infty J_{mp}(\lambda_1 r)a_{m}^iF_0\sin
mp\theta, \label{8}\\
H_r=& {\sum_{m=0}^\infty}\left[ \frac{2mp\epsilon_1
\omega}{\lambda_1^2r}J_{mp}(\lambda_1
r)a_{m}^i+\frac{2ik}{\lambda_1}J'_{mp}(\lambda_1
r)b_{m}^i\right]F_0\cos mp\theta, \label{9}\\
H_\theta=&-\sum_{m=1}^\infty \left[\frac{2\epsilon_1
\omega}{\lambda_1}J'_{mp}(\lambda_1
r)a_{m}^i+\frac{2ikmp}{\lambda_1^2r}J_{mp}(\lambda_1
r)b_{m}^i\right] F_0\sin mp\theta, \label{10}%\\
\end{align}
\begin{align}
H_z=&{2\sum_{m=0}^\infty}J_{mp}(\lambda_1 r)b_{m}^iF_0\cos
mp\theta. \label{11}
\end{align}
\end{subequations}
\end{widetext}
Here $k$ is the axial wave number, and $\lambda_1 $ is the
transverse wave number given by
\begin{equation}
\lambda_1^2=n_1^2\omega^2-k^2. \label{12}
\end{equation}
The $J_{mp}$'s are ordinary Bessel functions of order
$mp$, which are finite at the origin for $mp\ge0$, for $p$ non-integral, while
\begin{equation}
F_0=\exp(ikz-i\omega t) \label{13}
\end{equation}
is the $m=0$ version of the more general quantity
$F_m=\exp(imp\theta +ikz-i\omega t)$. %The primes on the summations
%in Eqs.~(\ref{7}), (\ref{9}) and (\ref{11}) mean that the $m=0$
%terms are taken with half weight. 
The expressions (\ref{6}) -
(\ref{11}) satisfy the electromagnetic boundary conditions on the
surfaces $\theta=0$ and $\theta=\alpha$ automatically, for
arbitrary values of the coefficients $a_{m}$ and $b_{m}$. 
The $i$ superscript on the coefficient refers to the interior
region.
%[There
%is one single restriction on the coefficients, namely $a_0=0$, in
%order to satisfy the conditions $E_z=0$ and $H_\theta=0$ on the
%surfaces. This restriction is already taken care of in the
%expansions (\ref{6}), (\ref{8}) and (\ref{10}), by letting the
%summations start from $m=1$.] 
The $a_{m}$ modes and the $b_{m}$
modes are independent of each other. 
%When writing the expansions
%above, we made use of the relation
%$J_{-mp}(x)=(-1)^{mp}J_{mp}(x)$.

Because of the closure of the region  at  $r=a$  the problem
becomes an eigenvalue problem. Only discrete values of the
transverse wave number $\lambda_1$ can occur. Let us distinguish
between the two kinds of modes:

(i) TM polarization (the $a_{m}^i$-modes), which correspond to
\begin{equation}
J_{mp}(\lambda_1 a)=0. \label{14}
\end{equation}
We denote the roots by $j_{mp,s}$, where $s=1,2,3....$ For a given
value of the axial wave number $k$ the energy eigenvalues are
accordingly
\begin{equation}
\omega_{msk}^{\rm TM}=\frac{1}{n_1a}\sqrt{j_{mp,s}^2+k^2a^2}, \quad m
\geq 1, \, s\geq 1. \label{15}
\end{equation}
(ii) TE polarization (the $b_{m}^i$-modes), which correspond to
zeroes of $J_{mp}'$,
\begin{equation}
\omega_{msk}^{\rm TE}=\frac{1}{n_1a}\sqrt{(j'_{mp,s})^2+k^2a^2}, \quad
m \geq 0, \, s\geq 1. \label{16}
\end{equation}
The interior zero-point energy per unit length is
\begin{equation}
\mathcal{E}^{\rm int}
=\frac12\int_{-\infty}^\infty \frac{dk}{2\pi}
\sum_{s=1}^\infty \left[\omega_{0sk}^{\rm TE}+\sum_{m=1}^\infty
(\omega_{msk}^{\rm TM}+\omega_{msk}^{\rm TE})\right]. \label{17}
\end{equation}
We here include the zero-point energy associated with the
azimuthally symmetric TE mode, although there is no such TM
mode.
%as we have seen, the
%coefficient $a_0=0$ always. 
%The conventional factor of 1/2 in front
%of the integral is absent because the sum over positive $m$'s
%already includes the contributions from both modes $m$ and $-m$ in
%the original mode sum running from $m=-\infty$ to $m=+\infty$.
%[THIS IS PROBABLY NOT TRUE---IS THERE A FACTOR OF 2 ERROR?]

 To simplify the formalism somewhat, we introduce the symbol
 $\mathcal{E}^{\rm int}_m$,
 \begin{equation}
 \mathcal{E}^{\rm int}_m=
\frac12\int_{-\infty}^\infty \frac{dk}{2\pi}\sum_{s=1}^\infty
 \left(\omega_{msk}^{\rm TM}+\omega_{msk}^{\rm TE}\right). \label{18}
 \end{equation}
For $m=0$ only the TE mode is to be included.
We now make use of the argument principle. Any meromorphic
function $g(\omega)$ of a complex variable $\omega$ satisfies the
equation
\begin{equation}
\frac{1}{2\pi i}\oint \omega \frac{d}{d\omega}\ln
g(\omega)d\omega=\sum \omega_{\rm zeros}-\sum \omega_{\rm poles},
\label{19}
\end{equation}
where $\omega_{\rm zeros}$ are the zeros and $\omega_{\rm poles}$  the
poles of $g(\omega)$ lying inside the integration contour. The
contour is chosen to be a large semi-circle in the right half-plane
with radius $R$,
closed by a straight line along the imaginary $z$ axis  from
$\omega=iR$ to $\omega=-iR$. A general advantage of this method is
that the multiplicities of zeros as well as for poles are
automatically taken care of.

In the present case it is evident that $g(\omega)$ can be chosen
as the product of $J_{mp}$ and $J'_{mp}$. There are no poles
involved, and the contribution of the large semi-circle goes to
zero when $R\rightarrow \infty$. Thus we obtain
\begin{align}
\mathcal{E}^{\rm int}_m=\frac12\frac{1}{2\pi i}\int_{-\infty}^\infty
&\frac{dk}{2\pi}\int_{i\infty}^{-i\infty}  d\omega\,\omega\notag \\
&\times\frac{d}{d\omega}\ln \left[J_{mp}(\lambda_1 a )J'_{mp}(\lambda_1
a)\right] . \label{20}
\end{align}
In the second integral, $k$ and $\omega$ are to be regarded as
independent variables in $\lambda_1=\lambda_1(k,\omega)$. We now
introduce the imaginary frequency $\zeta$ via $\omega \rightarrow
i\zeta$, whereby
\begin{equation}
 \lambda_1=\sqrt{n_1^2\omega^2-k^2}\rightarrow
 \sqrt{-(n_1^2\zeta^2+k^2)} \equiv i\kappa_1. \label{21}
 \end{equation}
 We thus encounter  Bessel functions of imaginary arguments,
 $J_{mp}(ix)$, with $x=\kappa_1 a,\, m\geq 0$. Introducing the
 modified Bessel function $I_\nu(x)$ via $J_\nu(ix)=i^\nu I_\nu(x)$ for
 arbitrary order $\nu$ we get
 \begin{equation}
 \mathcal{E}^{\rm int}_m=-\frac12\frac{1}{2\pi}\int_{-\infty}^\infty
 \frac{dk}{2\pi}\int_{-\infty}^\infty d\zeta\,\zeta \frac{d}{d\zeta}\ln
 \left[I_{mp}(x)I'_{mp}(x)\right]. \label{22}
 \end{equation}
Here we rewrite the derivative as
$d/d\zeta=(n_1^2a^2\zeta/x)d/dx$, and take into account the
symmetry properties $\int_{-\infty}^\infty dk \rightarrow
2\int_0^\infty dk$, $\int_{-\infty}^\infty d\zeta \rightarrow
2\int_0^\infty d\zeta $ to get
\begin{align}
\mathcal{E}_m^{\rm int}=-\frac{n_1^2a^2}{2\pi^2}\int_0^\infty &dk\int_0^\infty
\frac{\zeta^2 d\zeta}{x}\notag\\
&\times\frac{d}{dx}\ln
\left[I_{mp}(x)I'_{mp}(x)\right]. \label{23}
\end{align}
In the plane spanned by the axes $n_1\zeta$ and $k$ we may
introduce polar coordinates $X$ and $Y$
\begin{subequations}
\label{pc}
\begin{eqnarray}
X&=&n_1\zeta =\kappa_1 \cos \theta, \\
Y&=&k=\kappa_1 \sin \theta, \label{24}
\end{eqnarray}
\end{subequations}
fulfilling the relation $X^2+Y^2=\kappa_1^2$. The area element in
the $XY$ plane becomes $\kappa_1 d\kappa_1 d\theta=n_1d\zeta dk$.
The integration of the polar angle over the first quadrant then
becomes simple, $\int_0^{\pi/2} \cos^2\theta d\theta=\pi/4$, and we
get %when omitting the irrelevant factor $i^{2mp-1}$ inside the
%logarithm,
\begin{align}
\mathcal{E}_m^{\rm int}=&-\frac{1}{8\pi n_1a^2}\int_0^\infty dx\, x^2
\frac{d}{dx}\left[ I_{mp}(x)I'_{mp}(x)\right] \notag\\
=&-\frac{1}{8\pi n_1 a^2}\int_0^\infty dx\,x^2 \left[
\frac{I'_{mp}(x)}{I_{mp}(x)}+\frac{I''_{mp}(x)}{I'_{mp}(x)}
\right]. \label{25}
\end{align}
Going back to Eq.~(\ref{17}) we can thus write the interior
zero-point energy as
\begin{eqnarray}
\mathcal{E}^{\rm int}&=&-\frac{1}{8\pi n_1a^2}\bigg\{ {\sum_{m=1}^\infty}
\int_0^\infty x^2 dx  \left[
\frac{I'_{mp}(x)}{I_{mp}(x)}+\frac{I''_{mp}(x)}{I'_{mp}(x)}
\right] \nonumber\\
&&\quad\mbox{}+\int_0^\infty dx\,x^2\frac{I_0''(x)}{I_0'(x)}\bigg\},\label{26}
\end{eqnarray}
where the last term represents the TE $m=0$ mode.
No regularization procedure has been applied at this stage.

%%%%%%%%%%%%%%%%%%%%%%%%%%%%%%%%%%%%%%%%%%%%%%%%%%%%
%%%%%%%%%%%%%%%%% S E C T I O N %%%%%%%%%%%%%%%%%%%%
%%%%%%%%%%%%%%%%%%%%%%%%%%%%%%%%%%%%%%%%%%%%%%%%%%%%

\section{ Exterior region included, assuming perfectly conducting circular arc}\label{sec_ext}

We now include the exterior region $r \geq a$, still assuming the
circular arc at $r=a$ to be perfectly conducting.

A choice has to be made for what kind of medium to fill the space
$r>a$. One possible choice might be to assume a vacuum on the
outside. Another  natural choice would be to take the exterior
medium to be identical to the interior one. We will in this
section
 allow for a generalization of the last option, namely to
assume that the exterior space is filled with a medium with
arbitrary constants $\epsilon_2$ and $\mu_2$, but with the
restriction that their product is the same as in the interior:
\begin{equation}
\epsilon_2 \mu_2=\epsilon_1 \mu_1 =n^2. \label{27}
\end{equation}  We will refer to this situation as ``diaphanous.''
This condition implying the constancy of light everywhere has
under several occasions turned out to be convenient
mathematically, for instance in connection with the Casimir theory
for dielectric balls \cite{brevik82,brevik82b,brevik83,brevik84,brevik88,brevik89, kenneth02}, and in the Casimir theory
for the relativistic piecewise uniform string \cite{brevik90,brevik95b,li91,brevik94,brevik98b}  (a
review is given in Ref.~\cite{brevik02}). In the latter case, the
velocity of light is to be replaced with the velocity of sound.
The condition (\ref{27}) means in the present problem that
$\lambda$ takes the same value on the outside as on the inside
(assuming $k$ to take the same values on the two sides).  The principal
advantage of this assumption, which is not easily satisfied in
nature, is that in simple cases Casimir self-energies will turn out then to be
finite.

In the exterior region $r>a$ we have the expansions, keeping the
formalism at first quite general,
\begin{widetext}
\begin{subequations}
\begin{align}
E_r=&\sum_{m=1}^\infty \left[-\frac{2k}{\lambda_2}
{H_{mp}^{(1)}}'(\lambda_2r)a_{m}^e-\frac{2i\mu_2 \omega mp}{\lambda_2^2r}
{H_{mp}^{(1)}}(\lambda_2 r)b_{m}^e\right]F_0\sin mp\theta, \label{28}\\
E_\theta=&-{\sum_{m=0}^\infty}\left[ \frac{2kmp}{\lambda_2^2r}
{H_{mp}^{(1)}}(\lambda_2 r)a_{m}^e+\frac{2i\mu_2 \omega}{\lambda_2}
{H_{mp}^{(1)}}'(\lambda_2 r)b_{m}^e\right]F_0\cos mp\theta, \label{29}\\
E_z=&2i\sum_{m=1}^\infty {H_{mp}^{(1)}}(\lambda_2 r)a_{m}^e 
F_0\sin mp\theta, \label{30}\\
H_r=& {\sum_{m=0}^\infty}\left[ \frac{2mp\epsilon_2 \omega}{\lambda_2^2r}
{H_{mp}^{(1)}}(\lambda_2 r)a_{m}^e+\frac{2ik}{\lambda_2}
{H_{mp}^{(1)}}'(\lambda_2 r)b_{m}^e \right]F_0\cos mp\theta, \label{31}\\
H_\theta=&-\sum_{m=1}^\infty \left[\frac{2\epsilon_2 \omega}{\lambda_2}
{H_{mp}^{(1)}}'(\lambda_2 r)a_{m}^e+\frac{2ikmp}{\lambda_2^2r}
{H_{mp}^{(1)}}(\lambda_2 r)b_{m}^e\right] F_0\sin mp\theta, \label{32}\\
H_z=&{2\sum_{m=0}^\infty}H_{mp}^{(1)}(\lambda_2 r)b_{m}^eF_0\cos
mp\theta. \label{33}
\end{align}
\end{subequations}
\end{widetext}
As before, $F_0$ is given by Eq.~(\ref{13}). The presence of the
Hankel function of the first kind, $H_{mp}^{(1)}$, ensures proper
behavior (outgoing waves) at infinity.  The $e$ superscript refers
to exterior modes.

Let us now take into account the condition (\ref{27}), implying
$\lambda_1=\lambda_2 \equiv \lambda$, and consider the boundary
conditions. %Like above, we must have
% $a_0^e=0$ in order to satisfy the conditions
%$E_z=0$ and $H_\theta=0$ at the surfaces $\theta=0$ and
%$\theta=\alpha$ for arbitrary $r$. 
For the TM polarization (the
$a_{m}^e$-modes) we get
\begin{equation}
H_{mp}^{(1)}(\lambda a)=0, \quad m\geq 1, \label{34}
\end{equation}
whereas for the TE polarization (the $b_{m}^e$-modes),
\begin{equation}
{H_{mp}^{(1)}}'(\lambda a)=0, \quad m\geq 0. \label{35}
\end{equation}
The roots of these eigenvalue 
equations are complex---nevertheless, the argument
principle may be applied as has been explained in detail in many
places \cite{deraad81,bjm01,nest06}.
We can now calculate the exterior zero-point energy $\mathcal{E}^{\rm ext}$ in
the same way as above. The modified Bessel function $K_\nu$ is
introduced via $H_\nu^{(1)}(ix)=(2/\pi)i^{-(\nu+1)}K_\nu(x)$. For
the total zero-point energy/length $\mathcal{E}=
\mathcal{E}^{\rm int}+\mathcal{E}^{\rm ext}$ we  obtain
\begin{align}
\mathcal{ E}=-\frac{1}{8\pi na^2} &
\bigg\{{\sum_{m=1}^\infty}\int_0^\infty x^2 dx
\left[\frac{I_{mp}'(x)}{I_{mp}(x)}+\frac{I_{mp}''(x)}{I_{mp}'(x)}\right.
\notag \\
 &+\left.\frac{K_{mp}'(x)}{K_{mp}(x)}+\frac{K_{mp}''(x)}{K_{mp}'(x)}\right]
\notag\\
&+\int_0^\infty x^2dx\left[\frac{I''_0(x)}{I'_0(x)}+\frac{K''_0(x)}{K'_0(x)}
\right]\bigg\}
. \label{36}
\end{align}

We now must face up to the fact that our result contains an unremovable
divergence, associated with the nonzero $a_2$ heat kernel coefficient
found by Nesterenko et al.~\cite{nest01,nest03}.  This occurs precisely
because of the $m=0$ terms in Eq.~(\ref{36}).  If we were to write that
expression as
 \begin{eqnarray}
\mathcal{ E}&=&-\frac{1}{8\pi na^2} 
\bigg\{{\sum_{m=0}^\infty}{}'\int_0^\infty x^2 dx
\left[\frac{I_{mp}'(x)}{I_{mp}(x)}+\frac{I_{mp}''(x)}{I_{mp}'(x)}\right.
\nonumber \\
 &&\quad\mbox{}+\left.\frac{K_{mp}'(x)}{K_{mp}(x)}
+\frac{K_{mp}''(x)}{K_{mp}'(x)}\right]
\nonumber\\
&&\quad\mbox{}-\frac12
\int_0^\infty x^2dx\frac{d}{dx}\ln\left(\frac{I_0(x)}{I'_0(x)}
\frac{K_0(x)}{K_0'(x)}\right)\bigg\}\nonumber\\
&=&\tilde{\mathcal{E}}+\hat{\mathcal{E}}, \label{36a}
\end{eqnarray}
where the prime on the summation sign means that the $m=0$ terms
are counted with half weight,
we see in the following that the summation, $\tilde{\mathcal{E}}$,
 may now be rendered finite (see Appendix A),
but the residual correction, $\hat{\mathcal{E}}$, is divergent.

It is instructive to break up this residual zero mode contribution
into its Dirichlet (TM) and Neumann (TE) parts.  The former involves,
asympotically for large $x$
\begin{subequations}
\begin{equation}
I_0(x)K_0(x)\sim \frac1{2x}\left[1+\frac1{8x^2}+O\left(\frac1{x^4}\right)
\right],
\end{equation}
while the latter requires
\begin{equation}
I_0'(x)K_0'(x)\sim-\frac1{2x}\left[1-\frac3{8x^2}+O\left(\frac1{x^4}\right)
\right].
\end{equation}
\end{subequations}
Then the two contributions to the residual zero-mode terms are
\begin{subequations}
\begin{eqnarray}
&&-\frac12\int_0^\infty dx\,x^{2-s}\frac{d}{dx}\ln[I_0(x)K_0(x)]\nonumber\\
&&\quad\sim
\frac12\int_0^\infty dx\,x^{2-s}\left(\frac1x+\frac1{4x^3}+\dots\right),
\\
&&\frac12\int_0^\infty dx\,x^{2-s}\frac{d}{dx}\ln[I'_0(x)K'_0(x)]\nonumber\\
&&\quad\sim
\frac12\int_0^\infty dx\,x^{2-s}\left(-\frac1x+\frac3{4x^3}+\dots\right),
\end{eqnarray}
\end{subequations}
so the $1/x$ terms cancel between the two modes (alternatively those terms
may be removed by contact terms, as we will see in the following),
but the subleading $1/x^3$ terms constitute an unremovable logarthmic
divergence.  Here, we have indicated an analytic regularization by taking
$s$ to zero through positive values, which corresponds to the following
divergent terms as $s\to0$:
\begin{subequations}
\begin{eqnarray}
-\frac12\int_0^\infty dx\,x^{2-s}\frac{d}{dx}\ln
I_0(x)K_0(x)&\sim&\frac1{8s},
\\
\frac12\int_0^\infty dx\,x^{2-s}\frac{d}{dx}\ln
I'_0(x)K'_0(x)&\sim&\frac3{8s}.
\end{eqnarray}
\end{subequations}
These precisely correspond to the two mode contributions, adding up to
$1/2s$, found by Nesterenko et al.~\cite{nest01}.

This zero-mode divergence is due to the sharp corners
where the arc meets the wedge.  
We will proceed by setting this term aside,
and computing the balance of the Casimir free energy.  
We note there is a closely related problem
which Nesterenko et al.~\cite{nest03} dubbed a cone.  That is, we identify
the two wedge boundaries at $\theta=0$ and $\alpha$, and impose periodic
boundary conditions there.  This means that we may take the angular
function in the mode sums
to be $e^{imp\theta}$, where $m$ may be either positive or
negative, and where now $p=2\pi/\alpha$.  Now all modes, including the
zero modes ($m=0$) contribute equally, and the summation on $m$
becomes
\begin{equation}
\sum_{m=-\infty}^\infty=2\sum_{m=0}^\infty{}'
\end{equation}
with the zero modes both having 1/2 weight in the latter form.
(For the radial function in the interior we can only use $I_{|\nu|}$
in order that the solution be finite at the origin.)
Thus we get precisely $2\tilde{\mathcal{E}}$ [Eq.~(\ref{36a})] 
without the residual zero mode term $\hat{\mathcal{E}}$, 
and we have eliminated the unremovable logarithmic divergence.
This is because the sharp corners, where the arc meets the wedge,
have been removed because there is no wedge boundary.
So if the reader prefers, he or she may regard the rest of the 
discussion in this and the following section 
to refer to this situation, which will
introduce an additional factor of two into the Casimir free energy,
and with the restriction $p\ge1$, where $p=1$ corresponds to the
circular cylinder first considered in Ref.~\cite{deraad81}.

So in any case disregarding in the following the residual zero-mode pieces
$\hat{\mathcal{E}}$, we
consider now the regularization of the $\sum_{m=0}^\infty{}'$ terms
in Eq.~(\ref{36a}),  $\tilde{\mathcal{E}}$, which, in order
to be a Casimir energy,  ought to be given in such a form that it
reduces to zero in the limit when $a\rightarrow \infty$. 
This will eliminate the divergence associated with the apex,
which is not relevant to the force on the circular arc. It is
easy to satisfy this requirement by observing that for large
values of $x$, and for general $\nu$, we can approximate
\begin{equation}
I_\nu(x) \sim \frac{1}{\sqrt{2\pi x}}e^x, \quad
K_\nu(x)\sim \sqrt{\frac{\pi}{2x}}e^{-x}, \quad x\to \infty\label{37}
\end{equation}
implying that $I_\nu'\sim I_\nu$ and $K_\nu'\sim -K_\nu$.
Accordingly,
\begin{equation}
\frac{d}{dx}\ln \left(-I_\nu K_\nu I'_\nu K'_\nu \right)
\sim 2\frac{d}{dx}\ln (I_\nu K_\nu)\sim
-\frac{2}{x} \label{38}
\end{equation}
to leading order in $x$. This term is to be subtracted off from the integrand 
in Eq.~(\ref{36}).  The Casimir energy for the wedge becomes then
\begin{align}
\tilde{ \mathcal{E}}=-\frac{1}{8\pi na^2}
&{\sum_{m=0}^\infty}{}' \int_0^\infty x^2 dx
\left[\frac{I_{mp}'(x)}{I_{mp}(x)}+
\frac{I_{mp}''(x)}{I_{mp}'(x)}\right.\notag \\
&\left.+\frac{K_{mp}'(x)}{K_{mp}(x)}+
\frac{K_{mp}''(x)}{K_{mp}'(x)}+\frac{2}{x}\right]
. \label{39}
\end{align}
We may here perform a partial integration (the boundary terms at
$x=0$ and $x=\infty$ do not contribute),
\begin{align}
 \tilde{\mathcal{E}}=&\frac{1}{4\pi na^2} {\sum_{m=0}^\infty}{}'
\int_0^\infty xdx \notag \\
&\times\ln
\left[-4x^2I_{mp}(x)I_{mp}'(x)K_{mp}(x)K_{mp}'(x)\right] .
 \label{40}
\end{align}
It is helpful to introduce a quantity $\lambda_\nu(x)$ for
arbitrary order $\nu$,
\begin{equation}
\lambda_\nu(x)=(I_\nu(x) K_\nu(x))', \label{41}
\end{equation}
and to take into account the Wronskian $W\{I_\nu, K_\nu \}=-1/x$.
From this we calculate the following useful relationship
\begin{equation}
-4x^2I_\nu (x)I_\nu'(x)K_\nu(x)K_\nu'(x)=1-x^2\lambda_\nu^2(x),
\label{42}
\end{equation}
and so end up with the following convenient form for the Casimir
energy:
\begin{equation}
\tilde{\mathcal{E}}= \frac{1}{4\pi na^2}{\sum_{m=0}^\infty}{}'\int_0^\infty x\,dx \ln
\left[1-x^2\lambda_{mp}^2(x)\right].\label{43}
\end{equation}
This is thus the boundary-induced contribution to the zero-point
energy. If the boundary $r=a$ were removed and either the interior
or the exterior medium were chosen to fill the whole wedge
region, we would get $\tilde{\mathcal{E}}=0$. This is a property relying on the
condition (\ref{27}) above. The temperature is  assumed to be
zero.

Although the leading behavior of the Bessel functions has been
subtracted in Eq.~(\ref{43}), it is still not in general
finite.  We will see in the following section how a finite
self-energy may be extracted from this formula.  For now, we
observe that this is a generalization of the standard formal
result for a conducting circular cylinder, which is obtained
from this result in the special case $p=1$ \cite{deraad81}.   (The overall
$1/n$ comes from an elementary scaling argument \cite{brevik08}.)
When $p=1$ the expression (\ref{43}) is one-half that for a conducting
circular cylinder. Referring to the
perfectly conducting wedge boundaries, we see that the Casimir
energy for periodic boundary conditions, with period $2\pi$,
is twice the Casimir energy for a perfectly conducting boundary 
condition imposed
on the $\theta$ interval of $\pi$, a result obvious from
the replacement of $e^{im\theta}$ for $m$ of either sign in the former case by
$\sin m\theta$ or $\cos m\theta$, $m\ge0$, in the latter.  This
general observation, which is the theorem stated in 
Eq.~(2.49) of Ref.~\cite{miltonbook} (see also Ref.~\cite{AW}) 
will allow us to obtain numerical
results rather immediately.  For the periodic boundary condition
situation, which eliminates the zero-mode problem, $p=1$ is exactly
the circular cylinder problem, and there is no additional factor of
1/2.

%%%%%%%%%%%%%%%%%%%%%%%%%%%%%%%%%%%%%%%%%%%%%%%%%%%%
%%%%%%%%%%%%%%%%% S E C T I O N %%%%%%%%%%%%%%%%%%%%
%%%%%%%%%%%%%%%%%%%%%%%%%%%%%%%%%%%%%%%%%%%%%%%%%%%%

\section{Dielectric boundary at $r=a$}\label{sec_diel}

Assume now that the perfectly conducting arc at $r=a$ is removed
and replaced by a dielectric boundary, wherewith the interior and exterior
regions become coupled via electromagnetic
boundary conditions at $r=a$. As before, we  assume that the plane
surfaces $\theta=0$ and $\theta=\alpha$ are perfectly conducting
for all values of $r$. (Alternatively, we may impose periodic boundary
conditions there.)

\begin{figure}
  \includegraphics[width=2in]{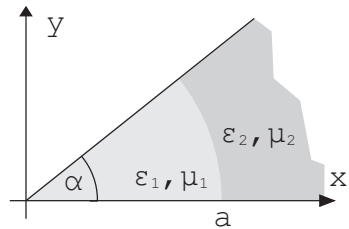}
  \caption{The wedge with a dielectric/diamagnetic boundary at $r=a$. 
In section \ref{sec_diel} we will allow $n_1\neq n_2$.}
  \label{fig_diel}
\end{figure}

We shall assume in the 
following that the media are arbitrary, with real and constant
parameters $\epsilon_1, \mu_1$ in the interior and
$\epsilon_2,\mu_2$ in the exterior, without any restriction
imposed on  their product.  This will, however, result in
general in a divergent Casimir self-energy.

 Let $\lambda_2$ be the transverse wave number in the exterior region,
 \begin{equation}
 \lambda_2^2=n_2^2\omega^2-k^2, \label{44}
 \end{equation}
 with $ n_2^2=\epsilon_2\mu_2$. The basic expansions are
 Eqs.~(\ref{6})-(\ref{11}) in the interior and
 Eqs.~(\ref{28})-(\ref{33}) in the exterior.

As for the boundary conditions at $r=a$, only the tangential field
components have to be taken into account.  From the continuity of
$E_z$ and $H_z$ we get respectively
\begin{equation}
J_{mp}(u)a_{m}^i=H_{mp}^{(1)}(v)a_{m}^e \label{45}
\end{equation}
and
\begin{equation}
J_{mp}(u)b_{m}^i=H_{mp}^{(1)}(v)b_{m}^e, \label{46}
\end{equation}
where we have defined
\begin{equation}
u=\lambda_1a, \quad v=\lambda_2 a. \label{47}
\end{equation}
\begin{widetext}
From the component $E_\theta$ we get
\begin{equation}
\frac{kmp}{u^2}J_{mp}(u)a_{m}^i+ \frac{i\mu_1
\omega}{u}J'_{mp}(u)b_{m}^i=
\frac{kmp}{v^2}H_{mp}^{(1)}(v)a_{m}^e+ \frac{i\mu_2\omega}{v}
{H_{mp}^{(1)}}' (v)b_{m}^e, \label{48}
\end{equation}
and from the component $H_\theta$,
\begin{equation}
\frac{i\epsilon_1 \omega}{u}
J'_{mp}(u)a_{m}^i-\frac{kmp}{u^2}J_{mp}(u)b_{m}^i
=\frac{i\epsilon_2\omega}{v} {H_{mp}^{(1)}}' (v)a_{m}^e
-\frac{kmp}{v^2}H_{mp}^{(1)}(v)b_{m}^e. \label{49}
\end{equation}
The two last equations mean that a superposition of the TM and  TE
waves is in general necessary to satisfy the boundary conditions.
The exception is the axially symmetric case $m=0$. The condition
for solution of the  set of linear equations is that the system
determinant vanishes. Observing the relation
\begin{equation}
u^2-v^2=(n_1^2-n_2^2)\omega^2a^2 \label{50}
\end{equation}
which follows from Eqs.~(\ref{44}) and (\ref{47}), we obtain after
some manipulations the condition
\begin{equation}
\left[
\frac{\mu_1}{u}\frac{J'_{mp}(u)}{J_{mp}(u)}-\frac{\mu_2}{v}
\frac{{H_{mp}^{(1)}}'(v)}{H_{mp}^{(1)}(v)}\right]
\left[\frac{\epsilon_1\omega^2}{u}\frac{J_{mp}'(u)}{J_{mp}(u)}-
\frac{\epsilon_2\omega^2}{v}\frac{{H_{mp}^{(1)}}'(v)}{H_{mp}^{(1)}(v)}\right]
=m^2p^2k^2\left(\frac{1}{v^2}-\frac{1}{u^2}\right)^2. \label{51}
\end{equation}
\end{widetext}
This is essentially the same transcendental eigenvalue equation as found for
 a step-index optical fiber (cf., for instance, Refs.~\cite{stratton41} 
or \cite{okamoto06}). In transmission problems, one is usually
interested in calculating the discrete values of the propagation
constant $k$, assuming that the waveguide is fed  with some
frequency $\omega$.  Here our intention is different, namely to
calculate the discrete values of $\omega$ on the basis of an input
value for the continuous axial wave vector $k$.  As we noted in the
previous section, this dispersion relation generalizes that for
a circular cylinder, the special case $p=1$.

It may be noted that the roots of Eq.~(\ref{51}) are both real and
complex.  Application of the argument principle to such a problem
is discussed in Ref.~\cite{nest08}.
%%%%%%%%%%%%%%%%%%%%% SUBSECTION %%%%%%%%%%%%%%%%%%%%%
\subsection{$n_1=n_2$}

The TE and TM modes decouple in the special case when 
$n_1=\sqrt{\epsilon_1\mu_1}=n_2=\sqrt{\epsilon_2\mu_2}$.
In this case, the dispersion relation (\ref{51}) reduces to
$\Delta\tilde\Delta=0$, where $\Delta$ and $\tilde \Delta$
are the two factors on the left-hand side of Eq.~(\ref{51}),
and then using the Wronskian, we find
after Euclidean rotation, $\omega\to i\zeta$,
\begin{equation}
\Delta\tilde\Delta=\frac{(\epsilon_1+\epsilon_2)}{4c^2
\epsilon_1\epsilon_2}\frac{1-\xi^2x^2(I_{mp}K_{mp})^{\prime2}}
{x^2I_{mp}^2(x)K_{mp}^2(x)}
\end{equation}
where $x=\kappa a$, $c=1/n$, and the reflection coefficient
(for either polarization)
\begin{equation}
\xi=\frac{\epsilon_2-\epsilon_1}{\epsilon_2+\epsilon_1}.
\end{equation}
We conclude that the formula for the (zero-mode subtracted) Casimir energy is
\begin{equation}
\tilde{\mathcal{E}}=\frac12\frac1{2\pi i}\int_{-\infty}^\infty\frac{dk}{2\pi}
\sum_{m=0}^\infty
{}'\int_{i\infty}^{-i\infty}d\omega\,\omega\frac{d}{d\omega}\ln g_m(x),
\label{zmsub}
\end{equation}
where
\begin{equation}
g_m(x)=1-\xi^2x^2\lambda_{mp}^{2},
\end{equation}
where $\lambda_{mp}$ is given by Eq.~(\ref{41}).
Here, we have again subtracted off the terms that would be present
if either medium filled the entire wedge.  (The divergence structure
of the zero-mode term subtracted from Eq.~(\ref{zmsub})  
is analyzed in Appendix B.)
Again cavalierly integrating by parts, we obtain, using the change of
variables (\ref{pc}),
\begin{equation}
\tilde{\mathcal{E}}=
\frac1{4\pi n a^2}\sum_{m=0}^\infty{}'\int_0^\infty
 dx \,x\ln[1-\xi^2x^2\lambda^2_{mp}].\label{eeee}
\end{equation}
  As expected, this
differs from the conducting case (\ref{43}) by the appearance of $\xi^2$ in
front of $\lambda_{mp}$.  The conducting case is obtained by setting $\xi=1$.
All of this is just as for the circular cylinder case, which is obtained
from the $p=1$ result by multiplying by a factor of 2.

Let us now extract both the $\xi=1$ (perfect conducting)
and the small $\xi$  results for arbitrary
$p$.  A simple route is to follow the method given in 
in \cite{milton99}
or in
Chap.~7 of Ref.~\cite{miltonbook}.
The point is simply that the uniform asymptotic expansion for the
modified Bessel functions yields an asymptotic expansion for large $p$.
Thus we can write (see the Appendix A for details)
\begin{equation}
2n\tilde{\mathcal{E}}=\frac{\xi^2}{16\pi a^2}\ln (2\pi/p)+\bar{\mathcal{E}}_0+
2\sum_{m=1}^\infty\bar {\mathcal{E}}_m,\label{sub}
\end{equation}
where
\begin{subequations}
\begin{eqnarray}
\bar{\mathcal{E}}_0&=&\frac1{4\pi a^2}\int_0^\infty dx\, x
\bigg[\ln\left(1-\xi^2x^2\lambda_0^2(x)\right)\nonumber\\
&&\qquad\mbox{}+\frac{\xi^2}4\frac{x^4}{(1+x^2)^3}\bigg],\\
\bar{\mathcal{E}}_m&=&\frac1{4\pi a^2}\int_0^\infty dx\, x
\bigg[\ln\left(1-\xi^2x^2\lambda_{mp}^2(x)\right)\nonumber\\
&&\qquad\mbox{}+\frac{\xi^2}4\frac{x^4}{(m^2p^2+x^2)^3}\bigg].
\end{eqnarray}
\end{subequations}
(Further details are given in the cited references.)  Because of the
subtractions in the integrals, they are convergent.  

\begin{figure}[tb]
  \includegraphics[width=3in]{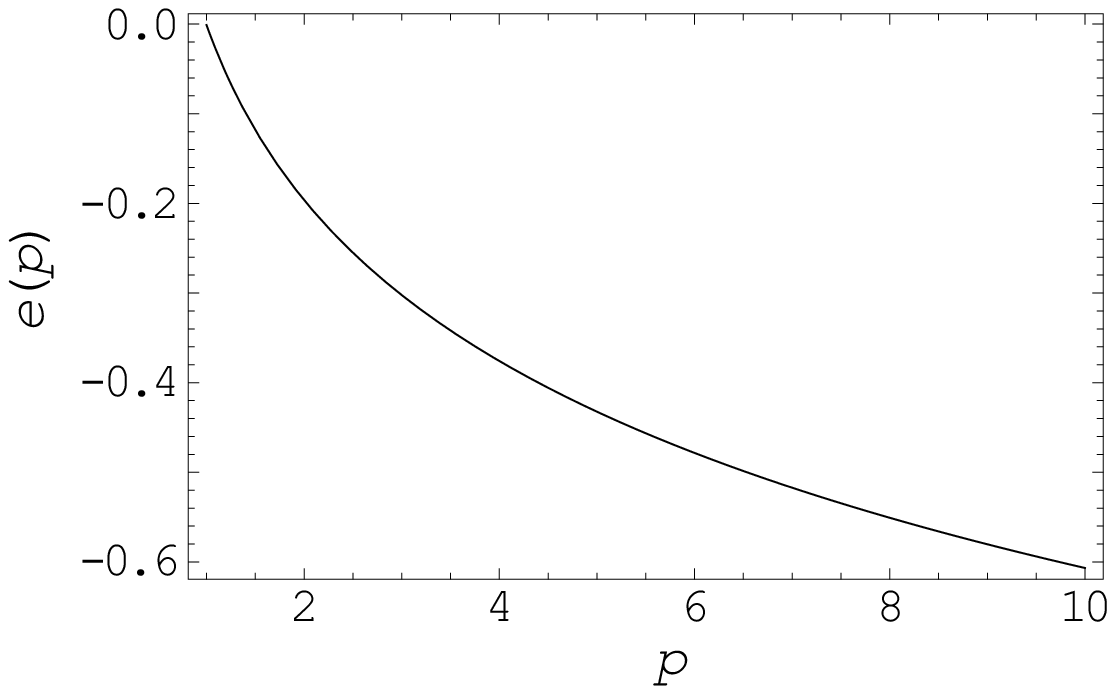} \\
  \includegraphics[width=3in]{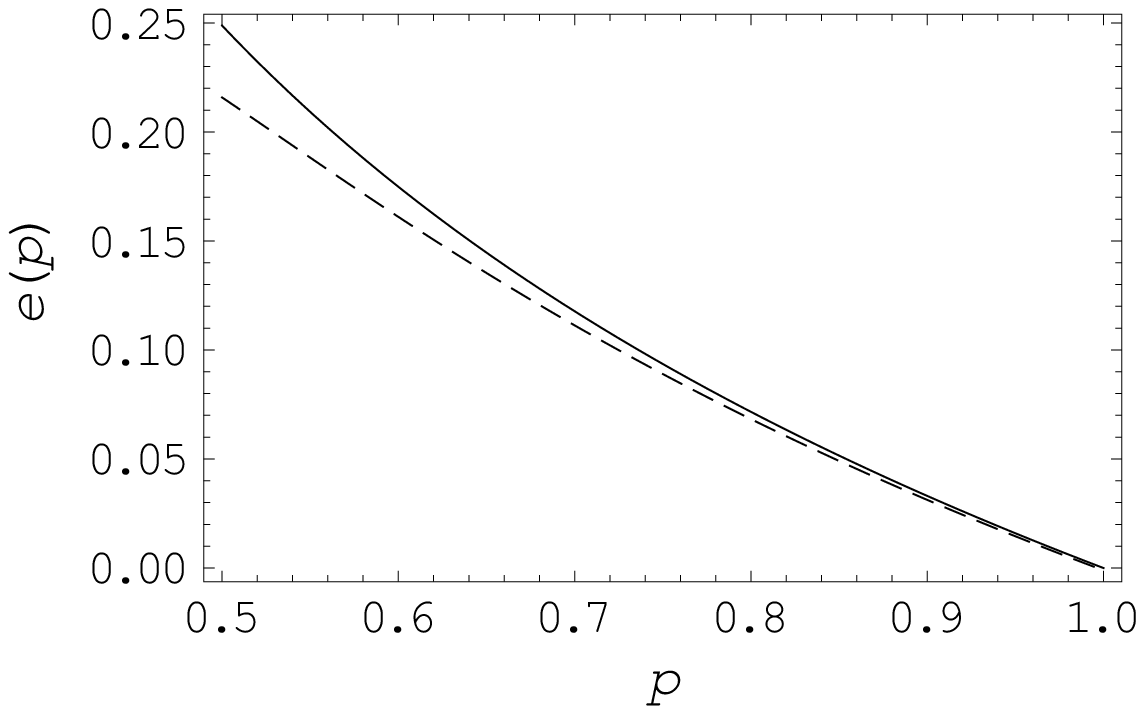}  
  \caption{Casimir energy for weak coupling, $\xi^2\ll1$,
as a function of $p$ which is related to the dihedral angle $\alpha=\pi/p$.
(\textbf{a}) $p>1$. (\textbf{b}) $0.5<p<1$;  the upper curve shows
the exact energy, the lower the leading asymptotic approximation,
obtained from Eq.~(\ref{weakdieldiam}) by setting $M=0$.}\label{fig1}
\end{figure}

Let us first consider $\xi$ as small, and keep only the terms
of order $\xi^2$.  Using the uniform
asymptotic approximants, we find for large $mp$,
\begin{equation}
\bar{\mathcal E}_m\sim\frac{\xi^2}{4\pi a^2}\left(\frac1{96m^2p^2}
-\frac7{3840 m^4p^4}+\dots\right),
\end{equation}
while numerical integration gives
\begin{equation}
\bar{\mathcal{E}}_0= \frac{\xi^2}{4\pi a^2}(-0.4908775).
\end{equation}
Thus 
\begin{eqnarray}
\tilde{\mathcal{E}}&=&
\frac{\xi^2}{8\pi n a^2}\bigg(-0.4908775+\frac14\ln2\pi/ p
+\frac{\pi^2}{288}\frac1{p^2}\nonumber\\
&&\quad\mbox{}-\frac{7\pi^4}{172800}\frac1{p^4}
+2\sum_{1}^M\left[f(mp)-g(mp)\right]\bigg)\nonumber\\
&\equiv&\frac{\xi^2}{8\pi n a^2}e(p),
\label{weakdieldiam}
\end{eqnarray}
where we have added and subtracted the first two terms in the
uniform asymptotic expansion,
\begin{equation}
g(\nu)=\frac1{96\nu^2}-\frac7{3840\nu^4},
\end{equation}
and $f$ is the integral appearing in $\bar{\mathcal{E}}_m$:
\begin{equation}
f(\nu)=\int_0^\infty dx\,x^3\left[-\lambda_\nu^2+\frac14\frac{x^2}
{(x^2+\nu^2)^3}\right].\label{fofnu}
\end{equation}
In principle we are to take the $M\to\infty$ limit in
Eq.~(\ref{weakdieldiam}).  In practice,
we may keep only a few terms in the $m$ sum.  For example, keeping
none of those corrections, that is setting $M=0$, we get for $p=1$,
$e(1)\approx -0.0010847$.  Keeping three terms is sufficient to
find that $e(1)$ is less than $10^{-6}$; indeed, the circular
cylinder value is $e(1)=0$ 
\cite{cavero05, cavero06, romeo05,romeo06,brevik07}.
This function $e(p)$ is plotted in Fig.~\ref{fig1}a, for $p>1$,
where it is sufficient to keep the leading asymptotic approximations;
for $p$ between 1/2 and 1 ($\alpha$ between $\pi$ and $2\pi$)
we must retain at least one correction,
$M=1$, as shown in Fig.~\ref{fig1}b.  (No observable change occurs
with larger $M$.)  Numerically, we see that the value for a cylinder
with a conducting septum ($p=1/2$) is indistinguishable from $e(0.5)=1/4$.
Recall for periodic boundary conditions on the wedge (the ``cone'')
$p=2\pi/\alpha\ge1$, and an additional factor of two appears in the
energy.

\begin{figure}[thb]
 \includegraphics[width=3in]{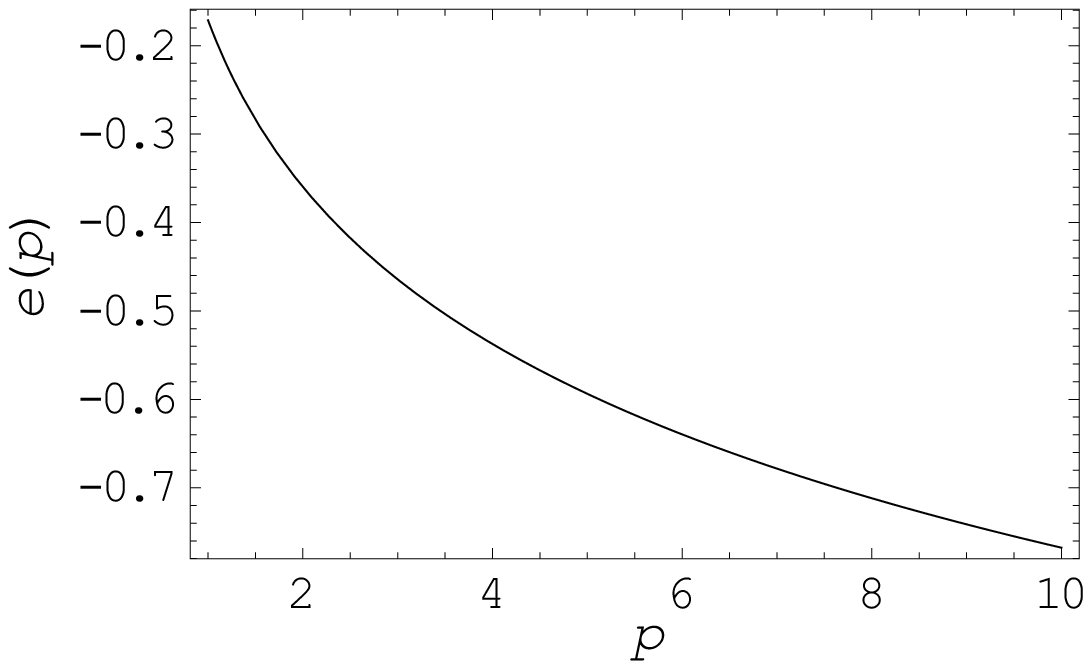} \\
 \includegraphics[width=3in]{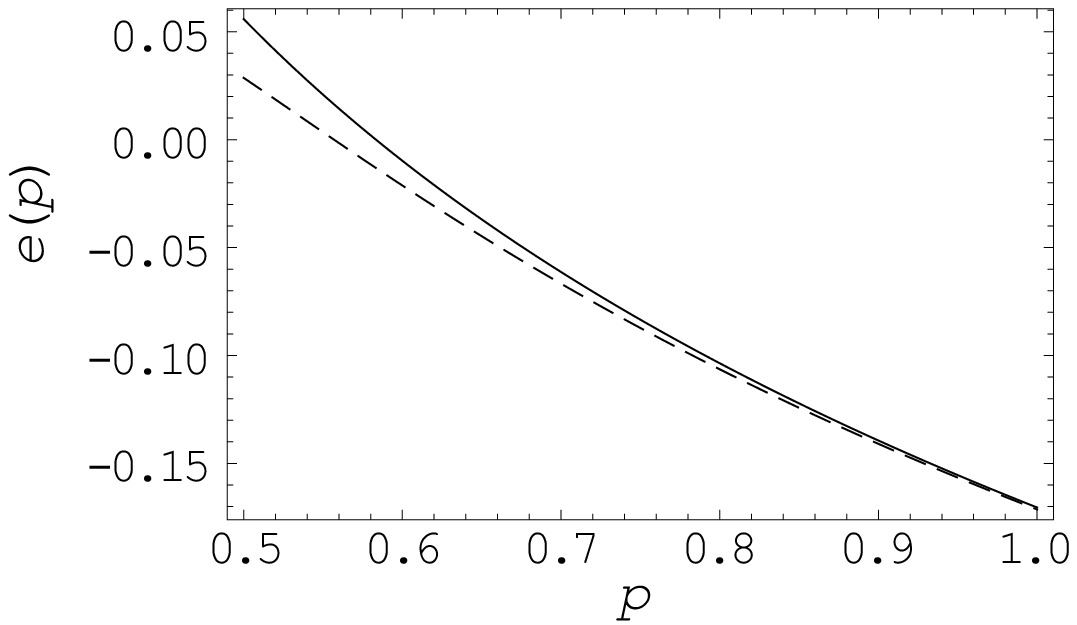}
  \caption{Casimir energy for strong coupling, $\xi^2=1$,
as a function of $p$, related to the dihedral angle by
$\alpha=\pi/p$. (\textbf{a}): $p>1$; in this region $M=0$ in Eq.~(\ref{stronge}) is
sufficient. (\textbf{b}): $0.5<p<1$; in this region $M=2$ in Eq.~(\ref{stronge}) is
sufficient, and the comparison with the $M=0$ result (lower curve) is made.}
  \label{fig2}
\end{figure}

Similarly, following the same references, we can obtain the
strong coupling (perfect conductor) limit, $\xi=1$.  This time the
formula for the energy is
\begin{equation}
\tilde{\mathcal{E}}=\frac1{8\pi n a^2}e(p),
\end{equation}
where 
\begin{eqnarray}
e(p)&=&-0.651752+\frac14\ln 2\pi/p
+\frac{7\pi^2}{2880}\frac1{p^2}-\frac{\pi^4}{32256}\frac1{p^4}
\nonumber\\
&&\qquad\mbox{}+2\sum_{m=1}^M\left[f(mp)-g(mp)\right],\label{stronge}
\end{eqnarray}
where again the limit $M\to\infty$ is understood.  Now $f$ is
given by
\begin{equation}
f(\nu)=\int_0^\infty dx\,x\left[\ln(1-x^2\lambda_\nu^2)+\frac14\frac{x^4}{
(x^2+\nu^2)^3}\right],
\end{equation}
and now the asymptotic terms are
\begin{equation}
g(\nu)=\frac7{960\nu^2}-\frac5{3584\nu^4}.
\end{equation}
Keeping no correction terms is 
already very good at $p=1$, where with $M=0$
$e(1)/(4\pi)  \approx-0.013633$, only slightly different from the exact
answer of $-0.01356$ \cite{deraad81}. Keeping just $M=1$ gives
exact coincidence to the indicated accuracy.  This function $e(p)$ is plotted
in Fig.~\ref{fig2}a for $p>0$ where the asymptotic approximation is
sufficient, while two correction terms are included in the region
$0.5<p<1$, as shown in Fig.~\ref{fig2}b.  It is curious that the
energy vanishes now not at $p=1$, but at $p=0.583$. (Again, recall only $p\ge1$ is relevant for periodic boundary conditions
on the wedge.)

%%%%%%%%%%%%%%%%%%%%% SUBSECTION %%%%%%%%%%%%%%%%%%%%%
\subsection{$n_1\ne n_2$, $\mu_1=\mu_2=1$}
\begin{figure}[tb]
  \includegraphics[width=3in]{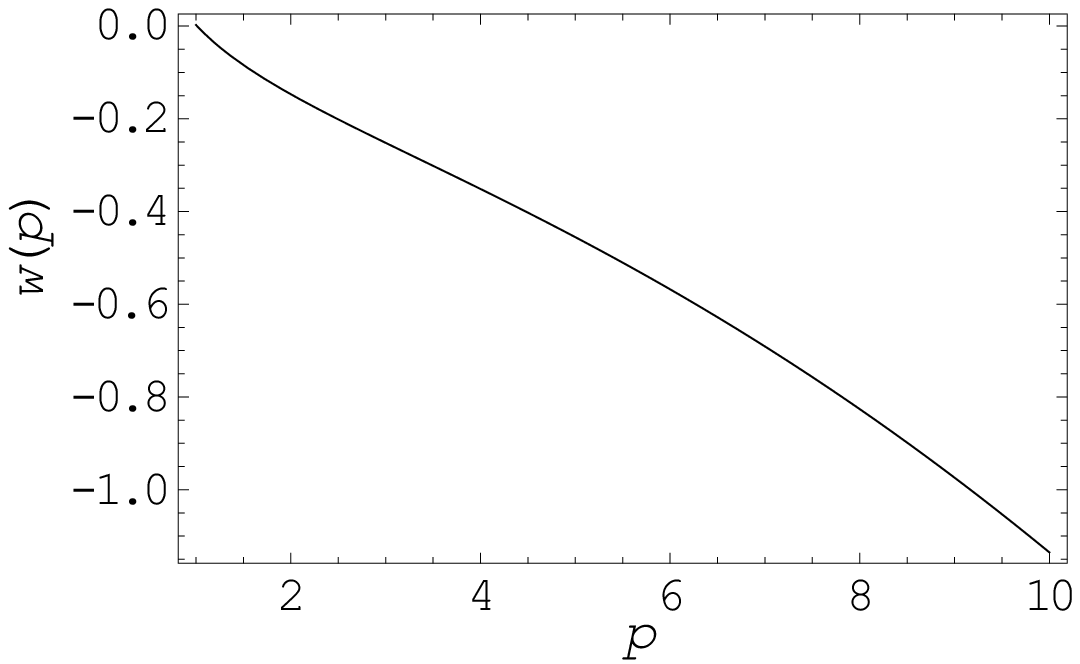} \\
 \includegraphics[width=3in]{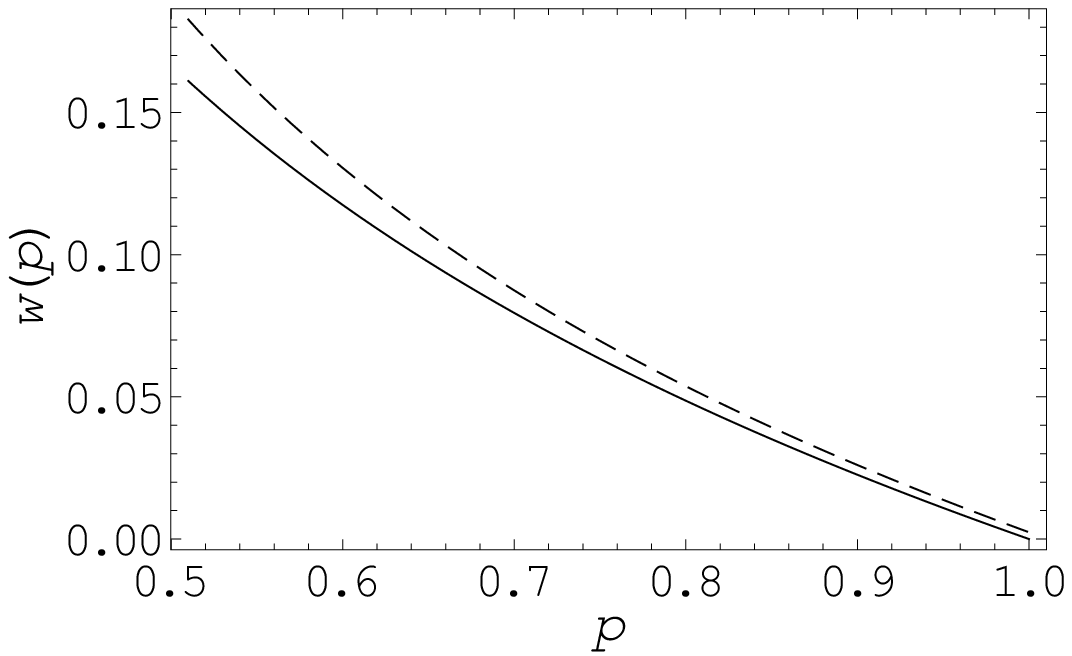}
  \caption{Casimir energy for weak coupling, $|\epsilon_1-\epsilon_2|\ll1$,
as a function of $p$, which is related to the dihedral angle 
$\alpha=\pi/p$, for a purely
dielectric wedge. (\textbf{a}): $p>1$; only the leading asymptotic terms have
been included. (\textbf{b}): $0.5<p<1$, 
 the effect of the remainder is significant; the upper curve
shows only the leading asymptotic terms, while the lower curve includes
the remainder function $r$.}
  \label{fig3}
\end{figure}

Finally, we can follow Ref.~\cite{cavero05}
to obtain the weak-coupling Casimir self-energy for a purely
dielectric wedge, where $\mu_1=\mu_2=1$.  We can only examine
the coefficient of $(\epsilon_1-\epsilon_2)^2$ because the result
is divergent in higher orders.  It is hardly necessary to give
details, since all that is necessary is to replace $m$ by $mp$
in the analysis given in that reference.  The energy per area
in the wedge is
\begin{equation}
\tilde{\mathcal{E}}=
\frac{(\epsilon_1-\epsilon_2)^2}{32\pi n a^2}\sum_{m=0}^\infty
{}'\int_0^\infty dy\,y^4 g_m(y),
\end{equation}
where the exact form of $g_m$ is elaborate, but has the asymptotic form
\begin{equation}
g_m(y)\sim\frac1{2m^2p^2}\sum_{k=1}^\infty \frac1{(mp)^k} f_k(z), \quad
mp\to\infty,
\end{equation}
where $y=mpz$, and $f_k$ are rational functions of $z$, given in
Ref.~\cite{cavero05}, about which all
we need to know here is
\begin{subequations}
\begin{eqnarray}
p^2\lim_{s\to0}\sum_{m=1}^\infty m^{2-s}\int_0^\infty dz\,z^{4-s}f_1(z)
&=&-p^2\frac{\zeta(3)}{16\pi^2},\nonumber\\
\\
\int_0^\infty dz\left(z^4 f_2(z)-\frac18\right)&=&0,\label{ct}\\
\lim_{s\to0}\sum_{m=0}^\infty{}'(mp)^{-s}\int_0^\infty dz \,z^{4-s}f_3(z)&=&
\frac5{32}\ln 2\pi/p,\nonumber\\ &&\label{4.27c}\\
\int_0^\infty dz\,z^4 f_4(z)&=&0,\label{log}\\
\frac1{p^2}\sum_{m=1}^\infty \frac1{m^2}\int_0^\infty dz\,z^4 f_5(z)&=&
\frac{19\pi^2}{7680}\frac1{p^2},\\
\int_0^\infty dz\,z^4 f_6(x)&=&0,\\
\frac1{p^4}\sum_{m=1}^\infty \frac1{m^4}\int_0^\infty dz\,z^4 f_7(z)&=&
-\frac{209\pi^4}{5806080}\frac1{p^4}.\nonumber\\
\end{eqnarray}
\end{subequations}
Here a contact term, which cannot contribute to any observable
force, has been removed from Eq.~(\ref{ct}).  Again, for the
precise definition of Eq.~(\ref{4.27c}) see Appendix A.
Then the Casimir energy per unit length of the dilute dielectric wedge is
\begin{equation}
\tilde{\mathcal{E}}\sim \frac{(\epsilon_1-\epsilon_2)^2}{64\pi n a^2}w(p),
\end{equation}
where
\begin{eqnarray}
w(p)&\approx&-p^2\frac{\zeta(3)}{16\pi^2}+\frac5{32}\ln (2\pi/p)
+\frac{19\pi^2}{7680p^2}-0.301590\nonumber\\
&&\qquad\mbox{}+\sum_{m=1}^4 r(mp)-\frac{0.000012}{p^2},
\end{eqnarray}
where 
\begin{equation}
r(\nu)=2\int_0^\infty dy\,y^4\left[g_\nu(y)-\frac1{2\nu^2}\sum_{k=1}^5
\frac1{\nu^k}f_k(y/\nu)\right],
\end{equation}
and we have used the next term in the asymptotic series to estimate
the contribution for $m\ge5$.
This, numerically, yields the correct
value of zero for $p=1$.
The values for the Casimir energy for larger values of $p$ are shown
in Fig.~\ref{fig3}a, and for smaller values of $p$ in Fig.~\ref{fig3}b.
For the septum case, the numerical value of $w(0.5)=0.1666$, 
which seems likely to represent exactly 1/6. (Periodic boundary
conditions restrict $p\ge1$.)

\section{Energy production in the sudden formation of a cosmic
string }

 As already mentioned, the electromagnetic theory of the wedge is
 related to the theory of cosmic strings. In general, cosmic
 strings are believed to be possible ingredients in the very early
 Universe; they are related to phase transitions. One particular aspect of
 this study is to estimate the energy production in the form of massless
 particles when a string is formed ``suddenly" at some instant
 $t=t_0$, where $t_0$ is a characteristic time  usually taken to be
 of order $10^{-40}$ s as is typical for grand unified theories
 (GUTs). One can calculate the number of particles associated with
 the formation of the string in terms of Bogoliubov coefficients
 relating the initial Minkowski metric to the (static) metric
 after the string has been formed. This approach was pioneered by
 Parker \cite{parker87} for massless scalar fields, assuming the
 string radius to be zero. Analogous calculations were made in
 Ref.~\cite{brevik95} in the electromagnetic case, still assuming
 the string radius to be zero, and in Ref.~\cite{brevik00} for the
 case of a finite string radius. (See also Ref.~\cite{khus99}.)

 In the following we shall investigate the following model: Let
 a cosmic string of vanishing radius and large length $L$ be formed suddenly
along  the cusp line of the wedge, i.e., along the $z$ axis. The
time of formation is  $t=t_0$. In the interior region we assume
that there is one single isotropic medium present, with refractive
index $n=\sqrt{\epsilon\mu}$. The interior region is closed by a
perfectly conducting arc with radius $r=a$. In the present
context, $a$ plays the role of a ``large" boundary; in connection
with  strings, $a$ is usually taken to be of the same order as $L$
\cite{parker87}. The finiteness of $a$ will moreover make it
 possible to normalize the fundamental modes in a
conventional way. As for the description of the electromagnetic
fields, we have to distinguish between the Minkowski metric
present for $t<t_0$ and the string metric for $t>t_0$.

\subsection{The case $t<t_0$}

We shall consider the TM mode only, for which the central field
component is $E_z$. For definiteness we reproduce the fields in
the $m,k$ mode here, in a convenient notation (replacing the
previous combination $2ia_m^i$  with the symbol  $N$):
\begin{equation}
E_z=NJ_{mp}(\lambda r)F_0\sin mp\theta, \label{5.1}
\end{equation}
\begin{equation}
E_r=\frac{ik}{\lambda}NJ'_{mp}(\lambda r)F_0\sin mp\theta,
\label{5.2}
\end{equation}
\begin{equation}
E_\theta=\frac{ikmp}{\lambda^2r}NJ_{mp}(\lambda r)F_0\cos
mp\theta, \label{5.3}
\end{equation}
\begin{equation}
H_z=0, \label{5.4}
\end{equation}
\begin{equation}
H_r=-\frac{imp\epsilon \omega}{\lambda^2r}NJ_{mp}(\lambda
r)F_0\cos mp\theta, \label{5.5}
\end{equation}
\begin{equation}
H_\theta=\frac{i\omega\epsilon}{\lambda}NJ'_{mp}(\lambda r)F_0\sin
mp\theta. \label{5.6}
\end{equation}
As before, $\lambda=\sqrt{n^2\omega^2-k^2}$, $F_0=\exp(ikz-i\omega
t)$, and $m \geq 1$.  The boundary condition on the arc is
$J_{mp}(\lambda a)=0$, giving the solutions $\lambda_{ms}$,
$s=1,2,3,...$ for the transverse wave number $\lambda$.

It is now convenient as an intermediate step to make use of the
formalism of scalar field theory. Define the scalar field mode
$\psi_{msk}$, satisfying Dirichlet boundary conditions on all
surfaces,  as
\begin{equation}\psi_{msk}=NJ_{mp}(\lambda_{ms}r)F_0\sin
mp\theta.\label{5.7}
\end{equation}
It is seen to have  the same form as the $m,s,k$ mode  of the
field component $E_z$. For reasons to become clear later, we
choose the magnitude $|N|$ of the normalization constant $N$ to be
\begin{equation}
|N|=\frac{1}{n}\sqrt{\frac{2}{\alpha\epsilon\omega_{msk}}}\,
\frac{\lambda_{ms}}{a|J_{mp+1}(\lambda_{ms}a)|}, \label{5.8}
\end{equation}
with $\omega_{msk}=(1/n)\sqrt{\lambda_{ms}^2+k^2}$.

We define the Klein-Gordon product as
\begin{equation}
(\psi_{msk},\psi_{m's'k'})=\frac{-i\epsilon
n^2}{\lambda_{ms}^2}\int
\psi_{msk}\stackrel{\leftrightarrow}{\partial_0}\psi^*_{m's'k'}
\,rdrd\theta dz, \label{5.9}
\end{equation}
and then get by direct calculation
\begin{equation}
(\psi_{msk},\psi_{m's'k'})=2\pi
\delta(k-k')\delta_{mm'}\delta_{ss'}. \label{5.10}
\end{equation}
Consider now the electromagnetic energy $W$ in the wedge region.
We may calculate it by integrating the energy density $w$ over the
volume:
\begin{equation}
W=\int w dV=\frac{1}{4}\int [\epsilon |E|^2+\mu |H|^2] rdrd\theta
dz, \label{5.11}
\end{equation}
using the general recursion equation $(J_\nu=J_\nu(x))$:
\begin{equation}
(J_\nu')^2+\frac{\nu^2}{x^2}J_\nu^2=\frac{1}{2}(J_{\nu-1}^2+J_{\nu+1}^2),
\label{5.12}
\end{equation}
as well as the integral formula
\begin{eqnarray}
&&\int_0^a \left[
J_{mp-1}^2(\lambda_{ms}r)+J_{mp+1}^2(\lambda_{ms}r)\right]rdr\nonumber\\
&&\qquad=a^2J_{mp+1}^2(\lambda_{ms}a),
\label{5.13}
\end{eqnarray}
which holds when $J_{mp}(\lambda_{ms}a)=0$. It is however simpler
to go via the axial energy flux $P$, given as
\begin{equation}
P=\int S_zdA, \label{5.14}
\end{equation}
where $dA=rdrd\theta$ is the cross-sectional area element, and
where
\begin{equation}
S_z=\frac{1}{2}\Re (E_rH_\theta^*-E_\theta H_r^*) \label{5.15}
\end{equation}
is the Poynting vector. As in any linear wave theory we can set
\cite{er}
\begin{equation}
P=\frac{W}{L}c_g, \label{5.16}
\end{equation}
where $c_g$ is the axial group velocity. From Eqs.~(\ref{5.14})
and (\ref{5.15}) we then get
\begin{equation}
P=\frac{\alpha \epsilon
ka^2\omega_{mks}}{8\lambda_{ms}^2}|N|^2J_{mp+1}^2(\lambda_{ms}a).
\label{5.17}
\end{equation}
In geometric units $P$  has the dimension $\rm cm^{-2}$. As
$c_g=d\omega/dk=k/(n^2\omega)$ we get for the energy per unit
length
\begin{equation}
\frac{W }{L}=\frac{\alpha \epsilon
n^2a^2\omega_{msk}^2}{8\lambda_{ms}^2}|N|^2J_{mp+1}^2(\lambda_{ms}a).
\label{5.18}
\end{equation}
We see that $W/L$ is expressible in terms of $|E_z|^2$ integrated
over the cross section:
\begin{equation}
\frac{W}{L}=\frac{\epsilon n^2\omega_{msk}^2}{2\lambda_{ms}^2}\int
|E_z|^2dA. \label{5.19}
\end{equation}
This relation will turn out to be useful in the following.

%\vspace{0.3cm}

 {\it Quantum theory.} We assume henceforth the real representation
 for the fields. The component $E_z({\bf r}, t) \equiv E_z(x)$,
 considered quantum-mechanically as a Hermitian operator, is
 expanded as
 \begin{equation}
 E_z(x)=\int_{-\infty}^\infty \frac{dk}{2\pi}\sum_{m=1}^\infty
 \sum_s
 \left[a_{msk}\psi_{msk}(x)+a^\dagger_{msk}\psi^*_{msk}(x)\right],
 \label{5.20}
 \end{equation}
 where $a_{msk}$ and $a^\dagger_{msk}$ are annihilation and
 creation operators satisfying the commutation relations
 \begin{equation}
 [a^{\vphantom{\dagger}}_{msk}, a^\dagger_{m's'k'}]=2\pi
 \delta(k-k')\delta_{mm'}\delta_{ss'}. \label{5.21}
 \end{equation}
 We now go back to the relation (\ref{5.19}), and require that the
 total energy $W$ associated with the $m,k,s$ mode is equal to the
 occupation number $\langle a^\dagger_{msk}
a^{\vphantom{\dagger}}_{msk}\rangle$ times
 the photon energy $\omega_{msk}$:
 \begin{equation}
 \frac{\epsilon n^2\omega_{msk}^2}{\lambda_{ms}^2}\int \langle
 E_z^2\rangle rdrd\theta dz=\langle a^\dagger_{msk}
a^{\vphantom{\dagger}}_{msk}+\frac12\rangle
 \omega_{msk}. \label{5.22}
 \end{equation}
 Here we insert the expansion (\ref{5.20}). Because of the
 orthogonality of (\ref{5.21}), the various modes decouple so that
 the total energy is a sum over the mode energies. For the mode
 $\psi_{msk}$, written in the form (\ref{5.7}), we get from the
 condition above the expression for the normalization constant
 $|N|$ already given in Eq.~(\ref{5.8}). If $n=1$ and
 $\alpha=2\pi$, we recover the expression given in
 Ref.~\cite{brevik95}.

 \subsection{The case $t>t_0$. The Bogoliubov transformation}

 After the sudden creation of the cosmic string along the cusp
 line (the $z$ axis) at the instant $t=t_0$, we assume that the
 string metric is static. All transient phenomena are intended to
 be taken care of via the use of the quantum mechanical sudden
 transformation below. We first have to establish the field
 expressions in the presence of the string metric. The central
 gravitational quantity appearing in the formalism will be
 \begin{equation}
 \beta=(1-4GM)^{-1}, \label{5.23}
 \end{equation}
 already introduced above in Eq.~(\ref{3}). In a string context, $\beta$ is
 believed to be very close to unity. Writing the field component
 $E_z$ as $E_z(r)\exp(ikz-i\omega t)\sin mp\theta$ we obtain the following
 equation for the quantity $E_z(r)$:
 \begin{equation}
 \left(
 \frac{d^2}{dr^2}
+\frac{1}{r}\frac{d}{dr}+\lambda^2-\frac{\beta^2m^2p^2}{r^2}\right)
 E_z(r)=0, \label{5.24}
 \end{equation}
 with $\lambda^2=n^2\omega^2-k^2$ as before. Introducing the
 symbol $\nu$ as
 \begin{equation}
 \nu=\beta m, \label{5.25}
 \end{equation}
 we can  write the fundamental $\nu,s,k$ mode as
 \begin{equation}
 \psi_{\nu sk}=N_\nu J_{\nu p}(\lambda_{\nu s}r)F_0\sin mp\theta,
 \label{5.26}
 \end{equation}
 with $F_0=\exp(ikz-i\omega_{\nu sk}t)$. The boundary condition on
 $r=a$ is  $J_{\nu p}(\lambda a)=0$, giving solutions
 $\lambda_{\nu s}$, $s=1,2,3...$ for the transverse wave number.

 The formalism now becomes quite similar to that given before in
 the non-gravitational case. 
We list the main formulas. The normalization constant $|N_\nu|$
 becomes
\begin{equation}
|N_\nu|=\frac{1}{n}\sqrt{\frac{2\beta}{\alpha\epsilon\omega_{\nu
sk}}}\, \frac{\lambda_{\nu s}}{a|J_{\nu p+1}(\lambda_{\nu s}a)|},
\label{5.27}
\end{equation}
and the Klein-Gordon product, defined as
\begin{equation}
(\psi_{\nu sk},\psi_{\nu' s'k'})=\frac{-i\epsilon n^2}{\beta
\lambda_{\nu s}^2}\int \psi_{\nu
sk}\stackrel{\leftrightarrow}{\partial_0}\psi^*_{\nu' s'k'}
\,rdrd\theta dz, \label{5.28}
\end{equation}
leads to
\begin{equation}
(\psi_{\nu sk},\psi_{\nu' s'k'})=2\pi \delta(k-k')\delta_{\nu
\nu'}\delta_{ss'}. \label{5.29}
\end{equation}
The quantum mechanical expansion for $E_z$ becomes
 \begin{equation}
 E_z(x)=\int_{-\infty}^\infty \frac{dk}{2\pi}\sum_{m=1}^\infty
 \sum_s
 \left[a_{\nu sk}\psi_{\nu sk}(x)+a^\dagger_{\nu sk}\psi^*_{\nu sk}(x)\right],
 \label{5.30}
 \end{equation}
with associated commutation relations
 \begin{equation}
 [a^{\vphantom{\dagger}}_{\nu sk}, a^\dagger_{\nu' s'k'}]=2\pi
 \delta(k-k')\delta_{\nu \nu'}\delta_{ss'}. \label{5.31}
 \end{equation}
 We have to specify the continuity conditions for the fields at
 the transition time $t_0$. The component $E_z$ will be required to be
 continuous,
 \begin{equation}
 E_z(x)\Big|_{t_0^-}=E_z(x)\Big|_{t_0^+}, \label{5.32}
 \end{equation}
\begin{widetext}
 as well as the Klein-Gordon product,
 \begin{equation}
 \frac{-i}{\lambda_{ms}^2}\int \left[ E_z
 \stackrel{\leftrightarrow}{\partial_0}E_z^*\right]_{t_0^-}rdrd\theta
 dz =\frac{-i}{\beta \lambda_{\nu s}^2}\int \left[[ E_z
 \stackrel{\leftrightarrow}{\partial_0}E_z^*\right]_{t_0^+}rdrd\theta dz,
 \label{5.33}
 \end{equation}
 from which we get
 \begin{equation}
 \left[\partial_0E_z(x)\right]_{t_0^-}=\frac{\lambda_{ms}^2}{\beta
 \lambda_{\nu s}^2}\left[\partial_0E_z(x)\right]_{t_0^+}.
 \label{5.34}
 \end{equation}
 {\it The Bogoliubov transformation}. We have now two kinds of
 basic modes, namely $\psi_{msk}$ for $t<t_0$, and $\psi_{\nu sk}$
 for $t>t_0$. There are correspondingly two vacuum states,
 satisfying the relations $a_{msk}|0\rangle_{msk}=0$ and $a_{\nu
 sk}|0\rangle_{\nu sk}=0$. As in Refs.~\cite{brevik95,brevik00} we
 may expand the modes in terms of each other:
 \begin{equation}
 \psi_{\nu sk}(x)=\int_{-\infty}^\infty
 \frac{dk'}{2\pi}\sum_{m's'}\left[\gamma(\nu
 sk|m's'k')\psi_{m's'k'}(x)+\delta(\nu sk|m's'k')\psi_{m's'k'}^*
 (x)\right], \label{5.35}
 \end{equation}
where $\gamma$ and $\delta$ are the Bogoliuobov coefficients
\cite{birrell82}. The corresponding expansions for the operators
are
\begin{equation}
a_{\nu sk}=\int_{-\infty}^\infty
\frac{dk'}{2\pi}\sum_{m's'}\left[\gamma(\nu
sk|m's'k')a_{m's'k'}+\delta^*(\nu sk|m's'k')a_{m's'k'}^\dagger
\right]. \label{5.36}
\end{equation}
It means that the average number of particles produced in the
$m,s,k$ mode per unit $k$ space interval becomes
\begin{equation}
\frac{dN_{msk}}{dk}=\int_{-\infty}^\infty
\frac{dk'}{2\pi}\sum_{m's'}|\delta(\nu sk|m's'k')|^2. \label{5.37}
\end{equation}
From Eq.~(\ref{5.35}) we obtain, when making use of the
normalization of the scalar product corresponding to string space,
\begin{equation}
 \delta(\nu sk|m's'k')=-(\psi_{\nu sk},\psi^*_{m's'k'})
=\frac{i\epsilon n^2}{\beta \lambda_{\nu s}^2}\int \psi_{\nu
sk}\stackrel{\leftrightarrow}{\partial_0}\psi_{m's'k'}\,rdrd\theta
dz. \label{5.38}
\end{equation}
Here we insert the expressions (\ref{5.26}) and (\ref{5.7}) for
$\psi_{\nu sk}$ and $\psi_{m's'k'}$, and for simplicity we put
$t_0=0$. Defining the quantity $I_{ss'}$ as
\begin{equation}
I_{ss'}=\frac{\int_0^a J_{\nu p}(\lambda_{\nu
s}r)J_{mp}(\lambda_{ms'}r)rdr}{a^2|J_{\nu p+1}(\lambda_{\nu
s}r)J_{mp+1}(\lambda_{ms'}r)|}, \label{5.39}
\end{equation}
we then obtain after some calculation
\begin{equation}
\delta(\nu
sk|m's'k')=-\frac{1}{\sqrt{\beta}}\frac{\lambda_{ms}}{\lambda_{\nu
s}}\,2\pi \delta(k+k') \delta_{mm'}
 \left[ \sqrt{\frac{\omega_{\nu
sk}}{\omega_{msk}}} -\sqrt{\frac{\omega_{msk}}{\omega_{\nu
sk}}}\right] I_{ss'}. \label{5.40}
\end{equation}
\end{widetext}
As the value of $\beta$ is very close to unity, we put $\beta=1$
everywhere except in the difference between the square roots. With
$J_{\nu p}\rightarrow J_{mp}$ and $\lambda_{\nu s'}\rightarrow
\lambda_{ms'}$, the numerator in (\ref{5.39}) reduces to
$(a^2/2)J^2_{mp+1}(\lambda_{ms}a)\delta_{ss'}$, so that
approximately
\begin{equation}
	   I_{ss'} = \frac{1}{2}\delta_{ss'}. \label{5.41}
\end{equation}
Moreover, by applying the integral operator $\int dk'/2\pi$ on
$[2\pi(k+k')]^2$ we obtain effectively the length $L$ of the
string. For the electromagnetic energy produced in the mode
$m,s,k$, per unit wave number interval, we then get
\begin{equation}
\frac{dW_{msk}}{dk}=\frac{\omega_{msk}}{L}\frac{dN_{msk}}{dk}
=\frac{1}{4}\omega_{msk}\left(
\frac{\omega_{\nu
sk}}{\omega_{msk}}+\frac{\omega_{msk}}{\omega_{\nu sk}}-2\right).
\label{5.42}
\end{equation}

There are two properties of this expression worth noticing:

\vspace{0.3cm}
(1) It is independent of the opening angle $\alpha$. The physical reason for 
this appears to be related to the fact that our region of quantization is the 
interior wedge region only.  All the produced energy is  taken to be channeled 
into the wedge region (we are thus not  cutting out a fraction  $\alpha/2\pi$ 
of the total produced energy).  This contrasts the behavior in the 
cylindrically symmetric case, where the produced energy is  azimuthally 
symmetric  in the whole region $0 < \theta < 2\pi $ \cite{parker87}.

(2)  The produced energy, when expressed in terms of frequencies,
does not contain the refractive index $n$ explicitly. Equation
(\ref{5.42}) is formally the same as Eq.~(52)  in
Ref.~\cite{brevik95}.

\vspace{0.3cm} We may process the expression further by making use
of the asymptotic formula for the roots of the Bessel function,
\begin{equation}
\lambda_{ms}a = s\pi +(m-\frac{1}{2})\frac{\pi}{2}. \label{5.43}
\end{equation}
Here it is of physical interest to consider the region around zero
axial wave number, $k\approx 0$. Then $\omega_{\nu sk}\rightarrow
\omega_{\nu s0}=\lambda_{\nu s}/n, \, \omega_{msk}\rightarrow
\omega_{ms0}=\lambda_{ms}/n$, leading to
\begin{equation}
 \sqrt{\frac{\omega_{\nu s0}}{\omega_{ms0}}}
-\sqrt{\frac{\omega_{ms0}}{\omega_{\nu
s0}}}=(\beta-1)\frac{m}{2s+m-\frac{1}{2}},
\label{5.44}
\end{equation}
where we have expanded in the small quantity $(\beta -1)$ to
second order.  Then,
\begin{equation}
\frac{dW_{msk}}{dk}\Big|_{k\approx
0}=\frac{\pi}{8na}(\beta-1)^2\frac{m^2}{2s+m-\frac{1}{2}}.
\label{5.45}
\end{equation}
We thus see that finally the factor $n$ turns up in the
denominator; this is a characteristic property of  Casimir energy
expressions for dielectrics \cite{brevik08}.

The simplest possibility $m=s=1$ yields
\begin{equation}
\frac{dW_{11k}}{dk}\Big|_{k\approx
0}=\frac{\pi}{20na}(\beta-1)^2=\frac{4\pi}{5na}(GM)^2.
\label{5.46}
\end{equation}
The total energy $W$ produced per unit length follows by
multiplying (\ref{5.46}) with the wave number width $\Delta k \sim
1/L\sim 1/a$ around $k=0$. We may take $a$ to be of the same order
as the horizon size $\sim t$, $t$ being the time  just after the
Big Bang. We thus get, when leaving $n$ unspecified,
\begin{equation}
W \sim \frac{1}{n}\left(\frac{GM}{t}\right)^2. \label{5.47}
\end{equation}
This is  a characteristic property of  cosmic string theory.

\section{Conclusions}
We have computed the Casimir free energy for a wedge-shaped
region bounded by perfectly conducting planes meeting in an
angle.  The wedge region is filled with an azimuthally symmetric
material which is discontinous at a radius $a$ from the intersection
axis.  In general the wedge geometry is plagued with divergence problems.
Familiar is the divergence associated with the apex, which is not
relevant to the force on the circular boundary.  But there are also
divergences associated with the corners where the circular arc meets
the wedge boundary.  These divergences are manifested only in the
$m=0$ modes, which possess no dependence on the angular coordinate,
and have here been isolated and disregarded in the calculational part
of this paper.  They will not be present if the perfectly conducting
boundary conditions on the wedge are replaced by periodic boundary
conditions, which restricts the parameter $p$ to be greater than unity.
Then, if the speed of light is the same both inside and outside
the radius $a$, the energy corresponding to changes in $a$ is
finite.  If the speed of light differs for $r<a$ and $r>a$, the
Casimir energy is finite only through second order in the 
discontinuity of the speed of light.  These results are seen
to be straightforward generalizations of results holding for
dielectric/diamagnetic circular cylinders, which are recovered
if $p=1$.
We also consider, in the ``sudden'' approximation, the electromagnetic
radiation produced by the appearance of a cosmic string in this
geometry.

\acknowledgments
The work of KAM was supported in part by grants from the US National
Science Foundation and the US Department of Energy.  He thanks
Jef Wagner and Prachi Parashar for helpful conversations.
We thank Steve Fulling for very helpful comments on the first
version of this paper, and especially an anonymous referee
for pointing out our originally incorrect treatment of the zero modes.

\appendix
\section{Analytic regularization of logarithmically divergent term}
The only subtlety in the numerical calculations in Sec.~\ref{sec_diel}
is how the superficially logarithmically divergent terms are regulated.
Starting from Eq.~(\ref{eeee}) we have
\begin{equation}
\tilde{\mathcal{E}}=\sum_{m=0}^\infty{}'\tilde{\mathcal{E}}_m, 
\end{equation}
where
\begin{subequations}
\begin{eqnarray}
n\tilde{\mathcal{E}}_0
&=&\bar{\mathcal{E}}_0-\frac{\xi^2}{4\pi a^2}\int_0^\infty
dx\frac{x^5}{4(1+x^2)^3},\\
n\tilde{\mathcal{E}}_m&=&\bar{\mathcal{E}}_m-\frac{\xi^2}{4\pi a^2}\int_0^\infty
dx\frac{x^5}{4(m^2p^2+x^2)^3}.
\end{eqnarray}
\end{subequations}
Here, it will be observed that the integrals over $x$ are logarithmically
divergent.  We will regulate them analytically by replacing in the numerator
of both $x^5\to x^{5-s}$, where we will at the end take $s$ to zero through
positive values.  Thus we have
\begin{eqnarray}
2n\tilde{\mathcal{E}}-
\bar{\mathcal{E}}_0-2\sum_{m=1}^\infty \bar{\mathcal{E}}_m
&=&-\frac{\xi^2}{16\pi a^2}\int_0^\infty dx\frac{x^{5-s}}{(1+x^2)^3}\nonumber\\
&&\times\left(1+2\sum_{m=1}^\infty(mp)^{-s}\right),
\end{eqnarray}
where we have let in the $m$ terms $x=mpz$.  Now the last factor is,
as $s\to 0$,
\begin{equation}
1+2\zeta(s)p^{-s}\to-s\left(\ln 2\pi-\ln p\right),
\end{equation}
while the integral diverges as $s\to0$:
\begin{equation}
\int_0^\infty dx\frac{x^{5-s}}{(1+x^2)^3}=\frac1s.
\end{equation}
The result (\ref{sub}) follows immediately. An identical argument
leads to Eq.~(\ref{4.27c}).  This argument demonstrates the importance
for achieving a finite result of including both TE and TM zero modes
with half weight.

\section{Convergence condition for additional energy term assuming high 
frequency transparency}

The energy expression (\ref{eeee}),
\begin{equation}
\tilde{\mathcal{E}} = \frac1{4\pi na^2}\sum_{m=0}^\infty {}^{{}^\prime} 
\int_0^\infty dxx\ln[1-\xi^2x^2\lambda_{mp}^2(x)]
\end{equation}
has an additional term consisting of one half the $m=0$ term for the TE mode 
minus one half times that of the TM mode. These modes are determined in the
diaphanous case by $\Delta\tilde\Delta=0$, where $\Delta$ and $\tilde\Delta$ 
are
the two factors in Eq.~(\ref{51}), which for the zero modes are proportional
to 
\begin{equation}
\Delta_0\tilde\Delta_0\propto\left(\frac1{\varepsilon_1}\frac{I'_0}{I_0}
-\frac1{\varepsilon_2}\frac{K'_0}{K_0}\right)
\left(\varepsilon_1\frac{I'_0}{I_0}-\varepsilon_2\frac{K'_0}{K_0}\right).
\end{equation}
Then, using the Wronskian, we see that the residual zero-mode term is
\begin{equation}
\hat{\mathcal{E}}
= -\frac1{16\pi na^2} \int_0^\infty dx\,x^2 \frac{d}{dx}\ln\frac{1+\xi x 
\lambda_0(x)}{1-\xi x \lambda_0(x)}.\label{b2}
\end{equation}
(This just says that the reflection coefficients for the two modes
are $\xi_{\rm TM}=\xi$ and $\xi_{\rm TE}=-\xi$.)

As in the perfectly conducting case, this is divergent, if $\xi$ is
constant, because
\be
(I_0(x)K_0(x))'\sim -\frac1{2x^2},\quad x\to \infty,
\ee
which means that the integral in Eq.~(\ref{b2}) is linearly divergent.
However, if $\xi$ is frequency dependent, so that
\be
\xi\sim \zeta^{-\beta},\quad\zeta\to\infty,
\ee
it is apparent that the integral becomes finite if $\beta>1$.


\begin{thebibliography}{99}
\bibitem{mostepanenko97}
V. M. Mostepanenko and N. N. Trunov, {\it The Casimir Effect and
Its Applications} (Oxford University Press, Oxford, 1997).

\bibitem{dowker78} J. S. Dowker and G. Kennedy, J. Phys. A {\bf 11}, 895 
(1978).
\bibitem{deutsch79} D. Deutsch and P. Candelas, Phys. Rev. D {\bf 20}, 
3063 (1979).
\bibitem{brevik96} I. Brevik and M. Lygren, Ann. Phys. {\bf 251}, 157 (1996).
\bibitem{brevik98} I. Brevik, M. Lygren and V. Marachevsky, Ann. Phys. 
{\bf 267}, 134 (1998).
\bibitem{brevik01} I. Brevik and K. Pettersen, Ann. Phys. {\bf 291}, 267 
(2001).
\bibitem{nesterenko02} V. V. Nesterenko, G. Lambiase and G. Scarpetta, 
Ann. Phys. {\bf 298}, 403 (2002)
\bibitem{nest01} V. V. Nesterenko, G. Lambiase, and G. Scarpetta,
J. Math. Phys. {\bf 42}, 1974 (2001).
\bibitem{nest03} V. V. Nesterenko, I. G. Pirozhenko, and J. Dittrich,
Class. Quantum Grav. {\bf 20}, 431 (2003).
\bibitem{rezaeian02}A. H. Rezaeian and A. A. Saharian, Class. Quant. Grav. 
{\bf 19}, 3625 (2002).
\bibitem{saharian07} A. A. Saharian, Eur. Phys. J. C {\bf 52}, 721 (2007).
\bibitem{saharian08}  A. A. Saharian, in {\it The Casimir Effect and Cosmology: A volume in honour of Professor Iver H. Brevik on the occasion of his 70th birthday} S. Odintsov et al. (eds.) (Tomsk State Pedagogical University Press, 2008), p.87, preprint {\it hep-th/0810.5207} .
\bibitem{mendes08} T. N. C. Mendes, F. S. S. Rosa, A. Ten\'{o}rio and 
C. Farina, J. Phys. A {\bf 41} 164029 (2008).
\bibitem{rosa08} F. S. S. Rosa, T. N. C. Mendes, A. Ten\'{o}rio and C. Farina,
 Phys. Rev. A {\bf 78} 012105 (2008).
\bibitem{sukenik93} C. I. Sukenik, M. G. Boshier, D. Cho, V. Sandoghdar 
and E. A. Hinds, Phys. Rev. Lett. {\bf 70} 560 (1993).

\bibitem{barton} G. Barton, Proc.\ R. Soc.\ London {\bf 410}, 175 (1987).

\bibitem{deraad81}
L. L. DeRaad, Jr. and K. A. Milton, Ann. Phys. {\bf 136}, 229
(1981).
\bibitem{milton99} K. A. Milton, A. V. Nesterenko and V. V. Nesterenko, Phys.
Rev. D {\bf 59}, 105009 (1999). 
\bibitem{godzinsky98} P. Gosdzinsky and A. Romeo, Phys.
Lett. B {\bf 441}, 265 (1998). 
\bibitem{lambiase99} G. Lambiase, V. V. Nesterenko and
M. Bordag, J. Math. Phys. {\bf 40}, 6254 (1999).

\bibitem{cavero05}
I. Cavero-Pel$\rm \acute{a}$ez and K. A. Milton, Ann. Phys. (N.Y.)
{\bf 320}, 108 (2005). 
\bibitem{cavero06} I. Cavero-Pel$\rm \acute{a}$ez and K. A.
Milton, J. Phys. A {\bf 39}, 6225 (2006). 
\bibitem{romeo05} A. Romeo and K. A. Milton, Phys. Lett. B {\bf 621}, 309 (2005)
\bibitem{romeo06} A. Romeo and K. A. Milton, J. Phys. A {\bf 39}, 6703 (2006)
\bibitem{brevik07} I. Brevik and A. Romeo, Physica Scripta {\bf 76}, 48 (2007).

\bibitem{vilenkin94}
A. Vilenkin and E. P. S. Shellard, {\it Cosmic Strings and other
Topological Defects} (Cambridge University Press, Oxford, 1994),
Sec.~7.
\bibitem{frolov87}
V. P. Frolov and E. M. Serebriany, Phys. Rev. D {\bf 35}, 3779
(1987).

\bibitem{khus99} N. R. Khusnutdinov and M. Bordag, Phys. Rev. D,
{\bf 59}, 064017 (1999).

\bibitem{bdm08} E. R. Bezzera de Mello, V. B. Bezerra, A. A. Saharian,
and A. S. Tarloyan, Phys. Rev. D {\bf 78}, 105007 (2008).

\bibitem{stratton41}
J. A. Stratton, {\it Electromagnetic Theory} (McGraw-Hill, New
York, 1941), p. 524.


\bibitem{brevik82} I. Brevik and H. Kolbenstvedt, Phys. Rev. D {\bf 25}, 1731 (1982).
\bibitem{brevik82b} I. Brevik and H. Kolbenstvedt, Ann. Phys. (N.Y.) {\bf 143}, 179 (1982).
\bibitem{brevik83} I. Brevik and H. Kolbenstvedt, Ann. Phys. (N.Y.){\bf 149}, 237 (1983).
\bibitem{brevik84} I. Brevik and H. Kolbenstvedt, Can. J. Phys. {\bf 62}, 805 (1984). 
\bibitem{brevik88} I. Brevik and G. Einevoll, Phys. Rev. D {\bf 37}, 2977 (1988). 
\bibitem{brevik89} I. Brevik and I. Clausen, Phys. Rev. D {\bf 39}, 603 (1989). 
\bibitem{kenneth02} O. Kenneth, I. Klich, A. Mann and M. Revzen, Phys. Rev. Lett. {\bf 89}, 033001 (2002).


\bibitem{brevik90} I. Brevik and H. B. Nielsen, Phys. Rev. D {\bf 41}, 1185 (1990).
\bibitem{brevik95b} I. Brevik and H. B. Nielsen, Phys. Rev. D {\bf 51}, 1869 (1995). 
\bibitem{li91} X. Li, X. Shi and J. Zhang, Phys. Rev. D {\bf 44}, 560 (1991).
\bibitem{brevik94} I. Brevik and E. Elizalde, Phys. Rev. D {\bf 49}, 5319 (1994). 
\bibitem{brevik98b} I. Brevik, A. A. Bytsenko and H. B. Nielsen, Class. Quant. Grav. {\bf 15}, 3383 (1998).

\bibitem{brevik02}
I. Brevik, A. A. Bytsenko and B. M. Pimentel, In {\it Theoretical
Physics 2002, Part 2}, Horizons in World Physics Volume 243
(Editors T. F. George and H. F. Arnoldus, Nova Science Publ, New
York, 2002), pp. 117-139.

\bibitem{bjm01} I. Brevik, B. Jensen, and K. A. Milton, Phys. Rev. D
{\bf 64}, 088701 (2001).

\bibitem{nest06} V. V. Nesterenko, J. Phys. A: Math. Gen. {\bf 39}, 6609
(2006).

 \bibitem{brevik08}
 I. Brevik and K. A. Milton, Phys. Rev. E {\bf 78}, 011124 (2008).

\bibitem{miltonbook} K. A. Milton, {\it The Casimir Effect}
(World Scientific, Singapore, 2001).
\bibitem{AW} J. Ambj\o rn and S. Wolfram, Ann. Phys. (N.Y.)
{\bf 147}, 1 (1983).
\bibitem{okamoto06}
K. Okamoto, {\it Fundamentals of Optical Waveguides}, 2nd ed.
(Elsevier, Amsterdam, 2000).
%\bibitem{abramowitz72}
%M. Abramowitz and I. A. Stegun, {\it Handbook of Mathematical
%Functions} (Dover, New York, 1972).

\bibitem{nest08} V. V. Nesterenko, J. Phys. A: Math. Theor. {\bf 41},
164005 (2008).


\bibitem{parker87}
 L. Parker, Phys. Rev. Lett. {\bf 59}, 1369 (1987).
 \bibitem{brevik95}
 I. Brevik and T. Toverud, Phys. Rev. D {\bf 51}, 691 (1995).
 \bibitem{brevik00}
 I. Brevik and A. G. Fr{\o}seth, Phys. Rev. D {\bf 61}, 085011
 (2000).

\bibitem{er} K. A. Milton and J. Schwinger, {\it Electromagnetic
Radiation: Variational Methods, Waveguides, and Accelerators}
(Springer, Berlin, 2006).
 \bibitem{birrell82}
 N. D. Birrell and P. C. W. Davies, {\it Quantum Fields in Curved
 Space} (Cambridge University Press, Cambridge, England, 1982).






\end{thebibliography}
\end{document}